\pdfoutput=1
\documentclass{article}

\usepackage[utf8]{inputenc} 
\usepackage[T1]{fontenc}    

\usepackage{amsmath} 
\usepackage{amsfonts} 
\usepackage[scale=1.0]{XCharter}

\usepackage[
  paper  = letterpaper,
  top    = 1.0in,
  bottom = 1.0in,
  ]{geometry}

\usepackage[dvipsnames,table]{xcolor}
\definecolor{shadecolor}{gray}{0.9}

\usepackage[final,expansion=alltext]{microtype}
\usepackage[english]{babel}
\usepackage[parfill]{parskip}
\usepackage{afterpage}
\usepackage{framed}

{\endMakeFramed}

\usepackage{lineno}

\usepackage{ragged2e}

\newcounter{parcount}

\usepackage{graphicx}
\usepackage{wrapfig}
\usepackage[labelfont=bf,font=footnotesize,width=.9\textwidth]{caption}
\usepackage[format=hang]{subcaption}

\usepackage{booktabs,multirow,multicol}       

\usepackage[algoruled,algo2e]{algorithm2e}
\usepackage{listings}
\usepackage{fancyvrb}
\fvset{fontsize=\normalsize}

\usepackage[colorlinks,linktoc=all]{hyperref}
\usepackage[all]{hypcap}
\hypersetup{citecolor=MidnightBlue}
\hypersetup{linkcolor=MidnightBlue}
\hypersetup{urlcolor=MidnightBlue}

\usepackage[nameinlink, capitalize]{cleveref}

\usepackage[acronym,smallcaps,nowarn]{glossaries}[=v4.46]

\lstdefinestyle{mystyle}{
    commentstyle=\color{OliveGreen},
    keywordstyle=\color{BurntOrange},
    numberstyle=\tiny\color{black!60},
    stringstyle=\color{MidnightBlue},
    basicstyle=\ttfamily,
    breakatwhitespace=false,
    breaklines=true,
    captionpos=b,
    keepspaces=true,
    numbers=left,
    numbersep=5pt,
    showspaces=false,
    showstringspaces=false,
    showtabs=false,
    tabsize=2
}
\usepackage{amsthm}

\usepackage{centernot}
\usepackage{nicefrac}       
\usepackage{mathtools}
\usepackage{amsbsy}
\usepackage{amstext}
\usepackage{thmtools}
\usepackage{thm-restate}

\begingroup
    \makeatletter
    \@for\theoremstyle:=definition,remark,plain\do{%
        \expandafter\g@addto@macro\csname th@\theoremstyle\endcsname{%
            \addtolength\thm@preskip\parskip
            }%
        }
\endgroup

\crefname{lemma}{lemma}{lemmas}
\Crefname{lemma}{Lemma}{Lemmas}
\crefname{thm}{theorem}{theorems}
\Crefname{thm}{Theorem}{Theorems}
\crefname{prop}{proposition}{propositions}
\Crefname{prop}{Proposition}{Propositions}
\crefname{assumption}{assumption}{assumptions}
\crefname{assumption}{Assumption}{Assumptions}

\usepackage{booktabs}
\makeatletter
\def\adl@drawiv#1#2#3{%
        \hskip.5\tabcolsep
        \xleaders#3{#2.5\@tempdimb #1{1}#2.5\@tempdimb}%
                #2\z@ plus1fil minus1fil\relax
        \hskip.5\tabcolsep}
\newcommand{\cdashlinelr}[1]{%
  \noalign{\vskip\aboverulesep
           \global\let\@dashdrawstore\adl@draw
           \global\let\adl@draw\adl@drawiv}
  \cdashline{#1}
  \noalign{\global\let\adl@draw\@dashdrawstore
           \vskip\belowrulesep}}
\makeatother

\renewcommand{\epsilon}{\varepsilon}

\declaretheorem[style=plain,name=Theorem]{theorem}

\declaretheorem[style=definition,sibling=theorem,name=Example]{example}

\newenvironment{example*}
 {\pushQED{\qed}\example}
 {\popQED\endexample}
\numberwithin{equation}{section}

\DeclarePairedDelimiterX\Set[1]{\lbrace}{\rbrace}%
{  #1 }

\usepackage[utf8]{inputenc} 
\usepackage[T1]{fontenc}    
\usepackage{hyperref}       
\usepackage{url}            
\usepackage{booktabs}       
\usepackage{amsfonts}       
\usepackage{nicefrac}       
\usepackage{microtype}      
\usepackage[table]{xcolor}  
\usepackage{amsmath}
\usepackage{amsthm}
\usepackage{wrapfig,lipsum,booktabs}
\usepackage{multirow}
\usepackage{subcaption}
\usepackage{algorithm}
\usepackage{algpseudocode}
\usepackage{hhline}
\usepackage{bbding}

\definecolor{mydarkblue}{rgb}{0,0.08,0.45}

\hypersetup{ %
    pdftitle={},
    pdfauthor={},
    pdfsubject={},
    pdfkeywords={},
    pdfborder=0 0 0,
    pdfpagemode=UseNone,
    colorlinks=true,
    linkcolor=mydarkblue,
    citecolor=mydarkblue,
    filecolor=mydarkblue,
    urlcolor=mydarkblue,
    pdfview=FitH
}

\usepackage{tcolorbox}
\tcbuselibrary{skins}
\usepackage{afterpage}
\definecolor{linen}{RGB}{250,240,230}
\usepackage{fontawesome5}
\definecolor{linen}{RGB}{250,240,230}
\usepackage{enumitem}
\usepackage{fontawesome5}
\usepackage{caption}
\usepackage{wrapfig,lipsum,booktabs}

\definecolor{FDColor}{RGB}{0,114,178}
\definecolor{FADColor}{RGB}{213,94,0}
\definecolor{OursRowColor}{RGB}{225,240,255}

\crefname{appsec}{appendix}{appendices}
\Crefname{appsec}{Appendix}{Appendices}

\usepackage{appendix}
\usepackage{array}
\newcolumntype{x}[1]{>{\centering\arraybackslash}p{#1pt}}
\newcolumntype{y}[1]{>{\raggedright\arraybackslash}p{#1pt}}

\usepackage{csquotes}
\usepackage[%
minnames=1,maxnames=99,maxcitenames=2,
style=alphabetic,
doi=false,
url=false,
giveninits=true,
hyperref,
natbib,
backend=bibtex,
sorting=nyt,
backref=true
]{biblatex}%
\renewbibmacro{in:}{%
  \ifentrytype{article}{}{\printtext{\bibstring{in}\intitlepunct}}}

\renewbibmacro*{journal}{%
  \iffieldundef{journaltitle}
    {}
    {\printtext[journaltitle]{%
       \printfield[noformat]{journaltitle}%
       \setunit{\subtitlepunct}%
       \printfield[noformat]{journalsubtitle}}}}

\DeclareFieldFormat{sentencecase}{\MakeSentenceCase*{#1}}

\renewbibmacro*{title}{%
  \ifthenelse{\iffieldundef{title}\AND\iffieldundef{subtitle}}
    {}
    {\ifthenelse{\ifentrytype{article}\OR\ifentrytype{inbook}%
      \OR\ifentrytype{incollection}\OR\ifentrytype{inproceedings}%
      \OR\ifentrytype{inreference}\OR\ifentrytype{misc}}
      {\printtext[title]{%
        \printfield[sentencecase]{title}%
        \setunit{\subtitlepunct}%
        \printfield[sentencecase]{subtitle}}}%
      {\printtext[title]{%
        \printfield[titlecase]{title}%
        \setunit{\subtitlepunct}%
        \printfield[titlecase]{subtitle}}}%
     \newunit}%
  \printfield{titleaddon}}

\AtEveryBibitem{%
\ifentrytype{article}{
    \clearfield{urldate}%
    \clearfield{eprint}
    \clearfield{eid}
}{}
\ifentrytype{book}{
    \clearfield{url}%
    \clearfield{urldate}%
    \clearfield{eprint}
}{}
\ifentrytype{collection}{
    \clearfield{url}%
    \clearfield{urldate}%
    \clearfield{eprint}
}{}
\ifentrytype{incollection}{
    \clearfield{url}%
    \clearfield{urldate}%
    \clearfield{eprint}
}{}
}

\AtEveryBibitem{
    \clearfield{pages}
    \clearfield{review}%
    \clearfield{series}
    \clearfield{volume}
    \clearfield{month}
    \clearfield{isbn}
    \clearfield{issn}
    \clearlist{location}
    \clearfield{series}
    \clearlist{publisher}
    \clearname{editor}
}{}

\usepackage{thm-restate}

\crefformat{equation}{(#2#1#3)}
\crefformat{figure}{Figure~#2#1#3}
\crefname{example}{Example}{Examples}
\crefname{lemma}{Lemma}{Lemmas}
\crefname{cor}{Corollary}{Corollaries}
\crefname{theorem}{Theorem}{Theorems}
\crefname{assumption}{Assumption}{Assumptions}

\usepackage{enumitem} 
\usepackage[separate-uncertainty=true,multi-part-units=single]{siunitx} 

\declaretheoremstyle[
spacebelow=\parsep,
    spaceabove=\parsep,
  mdframed={
    backgroundcolor=gray!10!white,     
    hidealllines=true, 
    innertopmargin=8pt, 
    innerbottommargin=4pt, 
    skipabove=8pt,
    skipbelow=10pt,
    nobreak=true
}
]{grayboxed}

\crefname{gassumption}{Assumption}{Assumptions}

\usepackage{xcolor}

\definecolor{WowColor}{rgb}{.75,0,.75}
\definecolor{SubtleColor}{rgb}{0,0,.50}

\newcounter{margincounter}

\usepackage[affil-it]{authblk}

\usepackage{makecell}

\usepackage{multirow}
\usepackage{tikz}
\usetikzlibrary{positioning}
\usetikzlibrary{arrows, automata}
\usetikzlibrary{patterns}

\def\showauthornotes{1}
\ifnum\showauthornotes=1
\newcommand{\Authornote}[2]{{\sf\small\color{blue}{[#1: #2]}}}
\else
\newcommand{\Authornote}[2]{}
\fi

\title{AudioGAR: Bridging Reconstruction and Generation in Latent Audio Generative Models}

\author{
  {\bfseries\upshape
  Xianghong Fang$^{1,*}$
  \quad
  Geeyang Tay$^{1,*}$
  \quad
  Wentao Ma$^{1}$
  } \\ \vspace{-5pt}
  {\bfseries\upshape
  Tim G. J. Rudner$^{1,2}$
  \quad
  Dehan Kong$^{1}$
  } \\
  {\upshape
  \vspace*{10pt}
  $^1$University of Toronto \quad
  $^2$Vijil \quad
  $^{*}$Equal contribution.
  } \\
  {\upshape
  \vspace*{15pt}
    \begin{center}
    \href{https://ml-maple-monk.github.io/Audio-GAR/}{\faGlobe\enspace Website}
    \quad
    \href{https://github.com/ml-maple-monk/Audio-GAR}{\faGithub\enspace Code \& Models}
    \end{center}
  }
}
\date{}

\begin{document}

\maketitle

\begin{abstract}
Latent audio generative models are typically trained in two stages: an audio codec is learned first, followed by a latent generative model. This decomposition leads to a decoder train-generation mismatch: the codec decoder is trained on encoder-induced latents but deployed on generator-produced latents at inference time. Across diverse datasets and latent generative models, we observe clear reconstruction-generation gaps under both FD and FAD, showing that strong reconstruction quality does not necessarily translate into strong end-to-end generation quality. A natural remedy is to adapt the decoder on generation-produced latents, but generated latents lack correspondence with source audio and therefore cannot directly provide the paired supervision used for decoder fine-tuning. We introduce \textbf{AudioGAR}, which constructs intermediate latents by perturbing encoder latents and denoising them through the frozen latent diffusion model. These latents form a trajectory from reconstruction toward generation, with lower-noise latents retaining source correspondence and supporting paired decoder fine-tuning. We fine-tune only the codec decoder on these latents, while keeping the codec encoder and latent generative model frozen. When applied to AudioX, AudioGAR substantially improves generative performance. It requires only 1.5\% of the original training audio hours and 0.26\% of the original training cost.
\end{abstract}


\section{Introduction}
\label{sec:introduction}
Recent audio and music generation models span autoregressive, masked-token, and diffusion-based approaches~\citep{Kreuk2023AudioGenTG,Copet2023SimpleAC,Garca2023VampNetMG,Ziv2024MaskedAG,Liu2023AudioLDMTG,Liu2024AudioLDM2L}. Although these models generate audio differently, they are typically trained in two stages. An audio codec first learns to compress waveforms into a lower-dimensional latent space and reconstruct them through its decoder. A latent generative model is then trained to generate within this learned latent space. This decomposition avoids the computational burden of generating directly in raw waveform space, and has been adopted by a wide range of audio and music generation systems~\citep{Liu2023AudioLDMTG,Liu2024AudioLDM2L,Ghosal2023TexttoAudioGU,Evans2025StableAO,Tian2026AudioXTurboAU}.

This two-stage training exposes the decoder to different latent distributions during training and generation. During codec training the decoder is optimized primarily on encoder-induced latents from raw audio, whereas at inference it is deployed on latents produced by the latent generative model. If these two latent distributions differ, strong reconstruction performance on encoder-induced latents need not carry over to generator-produced latents, a phenomenon we refer to as the \emph{decoder train-generation mismatch}. Recent work has begun to design audio codecs with downstream generative modeling in mind~\citep{Bralios2025ReBottleneckLR,Seetharaman2026TamingAV,Parker2026SAMEAS}, but the distribution shift at the decoder input has received little attention. The analogous mismatch has recently been studied in visual latent generative models~\citep{Fang2026GenerationAwareRecon}, but its presence and impact in latent audio generation have not been systematically examined.

We find clear evidence of this mismatch in latent audio generative models. Encoder-induced and generator-produced latents remain distinguishable in latent space on representative models and datasets (see Section~\ref{sec:distribution-mismatch}). Across a broader set of models and datasets, reconstruction and generation also exhibit substantial gaps under both Fr\'echet Distance (FD) and Fr\'echet Audio Distance (FAD) (see Table~\ref{tab:recon-gen-gap} in Section~\ref{sec:recon-gen-gap}). These results show that the mismatch is consequential for generation performance, motivating a dedicated approach to bridge this gap.

A natural way to address the decoder train-generation mismatch is to fine-tune the decoder on generation-relevant latents. However, fully generated latents lack correspondence with any source audio and therefore cannot directly provide the paired reconstruction supervision used for decoder fine-tuning. This raises a practical question:
\begin{tcolorbox}[
  enhanced,
  colback=blue!4,
  colframe=blue!55!black,
  leftrule=3pt,
  toprule=0.4pt,
  bottomrule=0.4pt,
  rightrule=0.4pt,
  arc=2pt,
  boxsep=1.5pt,
  left=8pt, right=8pt, top=4pt, bottom=4pt,
  width=0.90\linewidth,
  center,
  before skip=2pt,
  after skip=5pt,
]
\begin{center}
    {\bf Can we obtain generation-relevant latents that remain paired with the source audio for decoder fine-tuning?}
\end{center}
\end{tcolorbox}

We introduce \textbf{AudioGAR} for latent audio diffusion models. AudioGAR constructs intermediate latents between reconstruction and generation by perturbing encoder-induced latents with noise and denoising them through the frozen latent diffusion model. Increasing the noise level makes these latents more generation-relevant but gradually weakens their correspondence with the source audio. At lower noise levels, they retain correspondence with the source audio while already incorporating information from the generative process. We fine-tune only the codec decoder on these latents, keeping the encoder and latent generative model frozen. We term this fine-tuning strategy \emph{generation-aware codec decoder adaptation}.

\begin{table*}
\vspace{-12pt}
\centering
\caption{\small{\textbf{AudioGAR improves AudioX with less training cost and data.}
Relative cost and data are computed w.r.t. the original AudioX training. Training cost is reported as single-H100-equivalent wall-clock hours.}}
\vspace{-2.2ex}
\resizebox{1.0\textwidth}{!}{
\begin{tabular}{l|cc|cc|cc|cc}
\toprule
\multirow{2}{*}{Methods}
& \multicolumn{2}{c|}{Training Cost}
& \multicolumn{2}{c|}{Training Data}
& \multicolumn{2}{c|}{AudioCaps}
& \multicolumn{2}{c}{MusicCaps} \\
\hhline{~|--|--|--|--}
& Wall-Clock (Hours) & Rel. Cost
& Audio Hours & Rel. Data
& FD $\downarrow$ & FAD $\downarrow$
& FD $\downarrow$ & FAD $\downarrow$ \\
\hhline{~|--|--|--|--}
AudioX
& 4000 & --
& 33052.3 & --
& 11.83 & 1.61
& 9.56 & 1.60 \\
\textbf{AudioX + AudioGAR}
& \textbf{10.3} & \textbf{0.26\%}
& \textbf{508.3} & \textbf{1.5\%}
& \textbf{11.62} & \textbf{1.34}
& \textbf{8.33} & \textbf{1.10} \\
\bottomrule
\end{tabular}}
\vspace{-3.5ex}
\label{tab:audiox_audiogar_efficiency}
\end{table*}
We evaluate AudioGAR on AudioX~\citep{Tian2026AudioXTurboAU} across both audio and music generation tasks. On AudioCaps~\citep{Kim2019AudioCapsGC}, AudioGAR reduces FAD by $16.8\%$ and FD by $1.8\%$. On MusicCaps~\citep{Agostinelli2023MusicLMGM}, the gains are more pronounced, with FD reduced by $12.9\%$ and FAD by $31.3\%$. These improvements are achieved by fine-tuning only the codec decoder, using just $1.5\%$ of the audio hours and $0.26\%$ of the training cost of the original AudioX training (see Table~\ref{tab:audiox_audiogar_efficiency}). AudioGAR also generalizes to TangoMusic~\citep{Kong2024ImprovingTM}, reducing FD by 6.3\% and FAD by 20.5\% on music generation (see Table~\ref{tab:main_table}).

\vspace*{-2pt}
\textbf{Contributions.}
Our contributions are summarized as follows:
\vspace*{-2pt}
\begin{enumerate}[topsep=0pt, align=left, leftmargin=15pt, labelindent=1pt,
listparindent=\parindent, labelwidth=0pt, itemindent=!, itemsep=3pt, parsep=0pt]
\item We identify the \textbf{decoder train-generation mismatch} in latent audio generative models, showing that encoder-induced and generator-produced latents are distinguishable and that reconstruction and generation exhibit substantial performance gaps in music generation task.
\item We introduce \textbf{AudioGAR}, which constructs intermediate latents between reconstruction and generation distribution, with lower-noise latents retaining correspondence with the source audio.
\item We use AudioGAR latents for \textbf{generation-aware codec decoder adaptation}, fine-tuning only the codec decoder while keeping the encoder and latent generative model frozen. AudioGAR consistently improves FD and FAD across models and datasets while requiring only a small fraction of the original training audio and computational cost.
\end{enumerate}
\vspace{-8pt}
\section{Reconstruction and Generation in Latent Audio Models}
\label{sec:reconstruction-generation}
Latent audio generative models share a single codec decoder across two distinct pathways, reconstruction and generation, yet supply it with encoder-induced and generator-produced latents respectively. We examine this asymmetry from the perspective of the decoder's latent inputs by characterizing the two distributions, verifying their divergence empirically, and measuring the corresponding reconstruction-generation performance gap.

\subsection{Latent Distributions in Reconstruction and Generation}
\label{sec:latent-distribution}
Reconstruction and generation share the same codec decoder but use input latents from different sources. We formalize these two pathways and describe the corresponding evaluation metrics.
\vspace{-7pt}
\paragraph{Reconstruction.}
Given a waveform $x \sim p_{\text{data}}$, the codec encoder $\mathcal{E}_{\theta}$ produces a latent $z_e = \mathcal{E}_{\theta}(x)$, which the decoder $\mathcal{D}_{\phi}$ maps back to a waveform $\hat{x} = \mathcal{D}_{\phi}(z_e)$. We denote the encoder-induced latent distribution by $\mathcal{P}_e$. During codec training, the decoder is optimized to reconstruct source audio from encoder-induced latents.
\vspace{-7pt}
\paragraph{Generation.}
Given conditioning information $c$, the latent generative model $\mathcal{G}_{\varphi}$ produces a latent $z_g \sim p_{\varphi}(z|c)$, which is decoded by the same codec decoder as $\tilde{x} = \mathcal{D}_{\phi}(z_g)$. We denote the resulting generator-produced latent distribution by $\mathcal{P}_g$. The decoder is thus optimized on encoder-induced latents but deployed on generator-produced latents during generation. When $\mathcal{P}_e \neq \mathcal{P}_g$, its input distribution differs between these two pathways. We refer to this  distribution shift as the \textbf{decoder train-generation mismatch}.
\vspace{-5pt}
\paragraph{Evaluation metrics.}
We evaluate reconstructed and generated audio using Fr\'echet Distance (FD) with PANNs embeddings~\citep{Kong2019PANNsLP} and Fr\'echet Audio Distance (FAD; \citealp{kilgour19_interspeech}) with VGGish embeddings~\citep{Hershey2016CNNAF}. We use \textbf{rFD} and \textbf{rFAD} for reconstructed audio $\hat{x}$, and \textbf{gFD} and \textbf{gFAD} for generated audio $\tilde{x}$. Reconstruction metrics evaluate the codec's reconstruction pathway, with the decoder operating on encoder-induced latents $z_e \sim \mathcal{P}_e$. Generation metrics evaluate the full generative system, with the same decoder operating on generator-produced latents $z_g \sim \mathcal{P}_g$.

\subsection{Empirical Evidence of Distribution Mismatch}
\label{sec:distribution-mismatch}

\begin{wrapfigure}[19]{r}{0.65\textwidth}
\small
\vspace{-8.5mm}
\begin{center}
\includegraphics[width=0.65\textwidth]{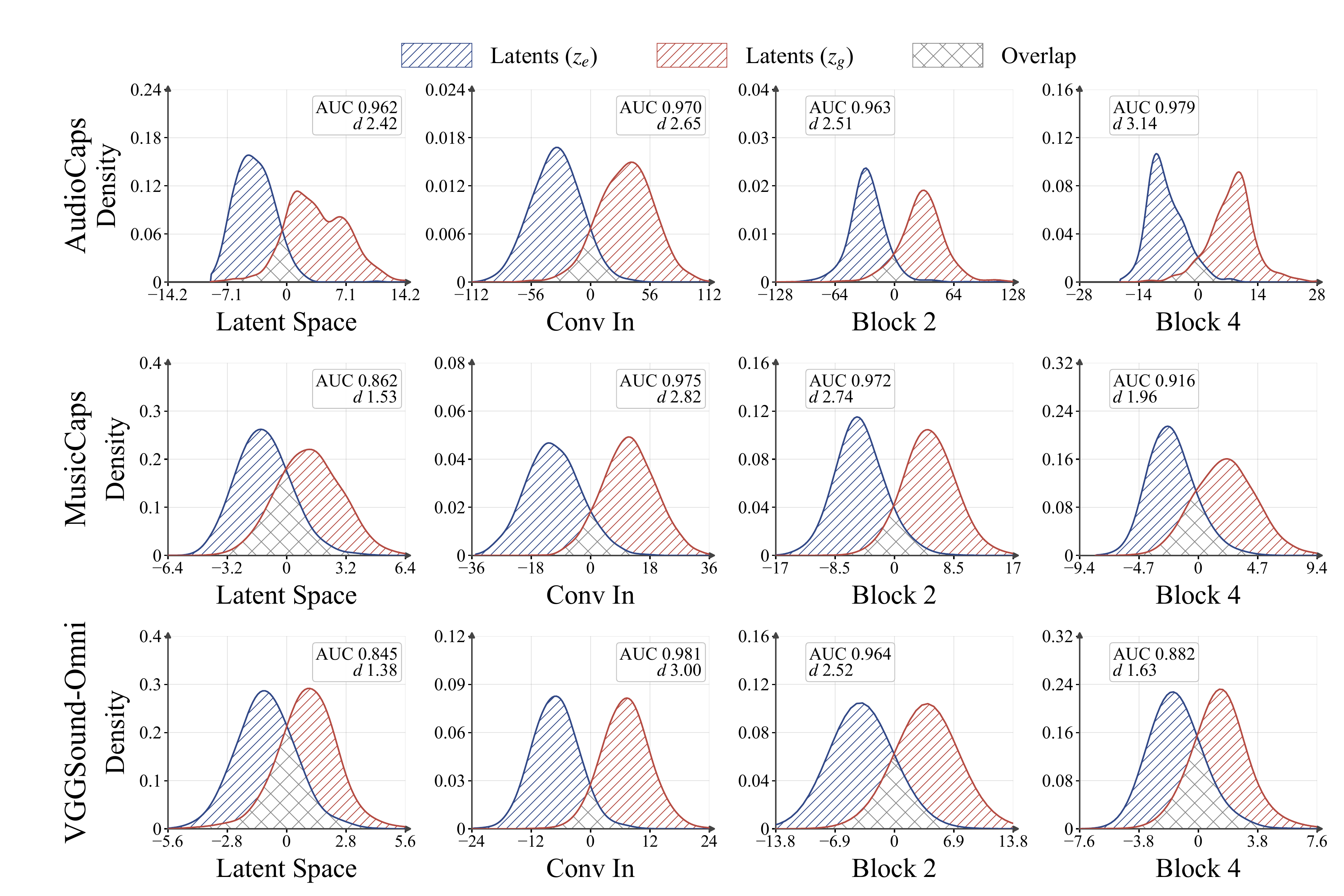}
\vspace{-7.5mm}
\caption{\small{LDA projections of $z_e$ and $z_g$ in AudioX. Rows correspond to AudioCaps, MusicCaps, and VGGSound-Omni datasets; columns show latent space and decoder features at Conv In, Block 2, and Block 4. Curves show Gaussian-smoothed, normalized histogram densities of the one-dimensional LDA scores.}}
\label{fig:lda-main}
\end{center}
\vspace{-2mm}
\end{wrapfigure}
To determine whether $\mathcal{P}_e$ and $\mathcal{P}_g$ differ in practice, we examine these two latent distributions in AudioX~\citep{Tian2026AudioXTurboAU} on AudioCaps~\citep{Kim2019AudioCapsGC}, MusicCaps~\citep{Agostinelli2023MusicLMGM},  and VGGSound-Omni~\citep{Dai2026Omni2SoundTU}. For each dataset, we obtain encoder-induced latents $z_e$ by encoding the real waveforms and generate $z_g$ using the corresponding text captions, so that the two latent collections are constructed from matched audio-caption examples. We then apply Linear Discriminant Analysis (LDA) to identify the direction that maximally separates $z_e$ and $z_g$, and visualize their one-dimensional projections using Gaussian-smoothed densities.

As shown in Figure~\ref{fig:lda-main}, encoder-induced and generator-produced latents are distinguishable in AudioX across all three datasets. Latent-space separation is strongest on AudioCaps, with an LDA AUC of 0.962, compared with 0.862 on MusicCaps and 0.845 on VGGSound-Omni. This distributional difference remains visible after the latents enter the decoder. Clear separation persists in the Conv In, Block 2, and Block 4 feature spaces, with particularly clear separation at Conv In and Block 2 for all three datasets. (see Appendix~\ref{app:decoder space} for feature extraction details). Figure~\ref{fig:lda-sao} in Appendix~\ref{app:additional-lda} shows the same qualitative pattern for Stable Audio Open~\citep{Evans2025StableAO} across all three datasets, with even clearer separation in the latent space. These results provide empirical evidence of a distribution mismatch, i.e., $\mathcal{P}_e \neq \mathcal{P}_g$, that occurs across the evaluated models and datasets and persists in intermediate decoder representations.  One possible explanation is that the learned latent generator $\mathcal{G}_{\varphi}$ cannot perfectly reproduce the encoder-induced latent distribution, resulting in the observed shift between encoder-induced and generator-produced latents.

\subsection{Reconstruction-Generation Performance Gap}
\label{sec:recon-gen-gap}

\begin{table*}[t]
\caption{\small{\textbf{Reconstruction-generation performance gap across latent audio generative models.} We report rFD and rFAD for reconstructed audio and gFD and gFAD for generated audio across multiple models on AudioCaps, MusicCaps, and VGGSound-Omni datasets. Reconstruction metrics are generally substantially lower than their generation counterparts. Lower values are better for all metrics.}}
\vspace{-2ex}
\label{tab:recon-gen-gap}
\centering
\resizebox{\textwidth}{!}{%
\begin{tabular}{
l
l|
>{\columncolor{FDColor!8}}c
>{\columncolor{FDColor!8}}c|
>{\columncolor{FADColor!8}}c
>{\columncolor{FADColor!8}}c
}
\toprule
Dataset & Models &  rFD ({\color{green!60!black}\bfseries$\pmb{\downarrow}$}) & gFD ({\color{green!60!black}\bfseries$\pmb{\downarrow}$}) & rFAD ({\color{green!60!black}\bfseries$\pmb{\downarrow}$}) & gFAD ({\color{green!60!black}\bfseries$\pmb{\downarrow}$}) \\
\hline
\multirow{10}{*}{\shortstack[c]{\centering AudioCaps}} &
\href{https://huggingface.co/haoheliu/audioldm2-full}{AudioLDM-2-Base}~\citep{Liu2024AudioLDM2L} & 3.19 & 11.54 & 1.31 & 2.04 \\
& \href{https://huggingface.co/haoheliu/audioldm2-full-large-1150k}{AudioLDM-2-Large}~\citep{Liu2024AudioLDM2L} & 3.20 & 11.53 & 1.31 & 1.86 \\
& \href{https://huggingface.co/haoheliu/AudioLDM-S-Full}{AudioLDM-S-Full}~\citep{Liu2023AudioLDMTG} & 3.05 & 21.85 & 1.27 & 4.79 \\
& \href{https://huggingface.co/HKUSTAudio/AudioX-MAF}{AudioX}~\citep{Tian2026AudioXTurboAU}  & 7.53 & 11.83 & 3.49 & 1.61 \\
& \href{https://huggingface.co/stabilityai/stable-audio-open-1.0}{Stable Audio Open}~\citep{Evans2025StableAO}  & 7.53 & 30.36 & 3.49 & 3.41 \\
& \href{https://huggingface.co/declare-lab/tango}{Tango}~\citep{Ghosal2023TexttoAudioGU}  & 2.97 & 14.47 & 1.14 & 1.52 \\
& \href{https://huggingface.co/declare-lab/tango-full}{Tango Full}~\citep{Ghosal2023TexttoAudioGU} & 2.97 & 22.23 & 1.14 & 4.10 \\
& \href{https://huggingface.co/declare-lab/tango-full-ft-audio-music-caps}{Tango Full (FT AudioCaps + MusicCaps)}~\citep{Ghosal2023TexttoAudioGU} & 2.97 & 9.77 & 1.14 & 2.59 \\
& \href{https://huggingface.co/declare-lab/tango-full-ft-audiocaps}{Tango Full (FT AudioCaps)}~\citep{Ghosal2023TexttoAudioGU} & 2.97 & 10.42 & 1.14 & 2.39 \\
& \href{https://huggingface.co/declare-lab/tango-af-ac-ft-ac}{Tango (AF-AC, FT AudioCaps)}~\citep{Kong2024ImprovingTM} & 2.97 & 12.08 & 1.14 & 2.55 \\
\hline
\multirow{8}{*}{\shortstack[c]{\centering MusicCaps}} & \href{https://huggingface.co/haoheliu/audioldm2-full-large-1150k}{AudioLDM-2-Large}~\citep{Liu2024AudioLDM2L} & 2.20 & 18.84 & 0.66 & 2.51 \\
& \href{https://huggingface.co/haoheliu/audioldm2-full}{AudioLDM-2-Base}~\citep{Liu2024AudioLDM2L} & 2.20 & 21.71 & 0.66 & 3.82 \\
& \href{https://huggingface.co/HKUSTAudio/AudioX-MAF}{AudioX}~\citep{Tian2026AudioXTurboAU} & 4.31 & 9.56 & 0.95 & 1.60 \\
& \href{https://huggingface.co/stabilityai/stable-audio-open-1.0}{Stable Audio Open}~\citep{Evans2025StableAO} & 4.31 & 37.72 & 0.95 & 3.42 \\
& \href{https://huggingface.co/declare-lab/tango}{Tango}~\citep{Ghosal2023TexttoAudioGU} & 2.18 & 46.80 & 0.56 & 5.64 \\
& \href{https://huggingface.co/declare-lab/tango-full}{Tango Full}~\citep{Ghosal2023TexttoAudioGU} & 2.18 & 35.67 & 0.56 & 5.78 \\
& \href{https://huggingface.co/declare-lab/tango-music-af-ft-mc}{TangoMusic}~\citep{Kong2024ImprovingTM} & 2.04 & 15.20 & 0.42 & 1.85 \\
& \href{https://huggingface.co/declare-lab/tango-full-ft-audio-music-caps}{Tango Full (FT AudioCaps + MusicCaps)}~\citep{Ghosal2023TexttoAudioGU} & 2.18 & 20.43 & 0.56 & 3.61 \\
\hline
\multirow{11}{*}{\shortstack[c]{\centering VGGSound-Omni}} & \href{https://huggingface.co/haoheliu/audioldm2-full}{AudioLDM-2-Base}~\citep{Liu2024AudioLDM2L} & 1.64 & 10.03 & 0.49 & 1.72 \\
& \href{https://huggingface.co/haoheliu/audioldm2-full-large-1150k}{AudioLDM-2-Large}~\citep{Liu2024AudioLDM2L} & 1.64 & 9.93 & 0.49 & 1.49 \\
& \href{https://huggingface.co/haoheliu/AudioLDM-S-Full}{AudioLDM-S-Full}~\citep{Liu2023AudioLDMTG} & 1.63 & 18.78 & 0.50 & 3.44 \\
& \href{https://huggingface.co/HKUSTAudio/AudioX-MAF}{AudioX}~\citep{Tian2026AudioXTurboAU} & 3.66 & 9.67 & 1.91 & 1.74 \\
& \href{https://huggingface.co/stabilityai/stable-audio-open-1.0}{Stable Audio Open}~\citep{Evans2025StableAO} & 3.66 & 26.07 & 1.91 & 2.68 \\
& \href{https://huggingface.co/declare-lab/tango}{Tango}~\citep{Ghosal2023TexttoAudioGU} & 1.62 & 24.03 & 0.50 & 2.88 \\
& \href{https://huggingface.co/declare-lab/tango-full}{Tango Full}~\citep{Ghosal2023TexttoAudioGU} & 1.62 & 19.01 & 0.50 & 3.69 \\
& \href{https://huggingface.co/declare-lab/tango-music-af-ft-mc}{TangoMusic}~\citep{Kong2024ImprovingTM} & 1.62 & 13.43 & 0.50 & 2.36 \\
& \href{https://huggingface.co/declare-lab/tango-af-ac-ft-ac}{Tango (AF-AC, FT AudioCaps)}~\citep{Kong2024ImprovingTM} & 1.62 & 11.72 & 0.50 & 1.46 \\
& \href{https://huggingface.co/declare-lab/tango-full-ft-audio-music-caps}{Tango Full (FT AudioCaps + MusicCaps)}~\citep{Ghosal2023TexttoAudioGU} & 1.62 & 13.26 & 0.50 & 2.28 \\
& \href{https://huggingface.co/declare-lab/tango-full-ft-audiocaps}{Tango Full (FT AudioCaps)}~\citep{Ghosal2023TexttoAudioGU} & 1.62 & 19.12 & 0.50 & 3.06 \\
\bottomrule
\end{tabular}
}
\vspace{-3ex}
\end{table*}

We next examine whether the latent distribution mismatch identified in Section~\ref{sec:distribution-mismatch} translates into a measurable performance gap at the decoder output. Table~\ref{tab:recon-gen-gap} compares distribution-level reconstruction performance, measured by rFD and rFAD, with distribution-level generation performance, measured by gFD and gFAD, across 10 latent generative models on AudioCaps~\citep{Kim2019AudioCapsGC}, 8 on MusicCaps~\citep{Agostinelli2023MusicLMGM}, and 11 on VGGSound-Omni~\citep{Dai2026Omni2SoundTU}. In every case, rFD is substantially lower than gFD, and rFAD is also lower than gFAD in 26 of the 29 evaluated model-dataset combinations, indicating that strong decoder performance on encoder-induced latents does not reliably transfer when the same decoder receives generator-produced latents. The gap can be substantial: on MusicCaps, for example, Tango yields rFD = 2.18 against gFD = 46.80.

\begin{wrapfigure}[12]{r}{0.50\textwidth}
\small
\vspace{-2mm}
\begin{center}
\includegraphics[width=0.50\textwidth]{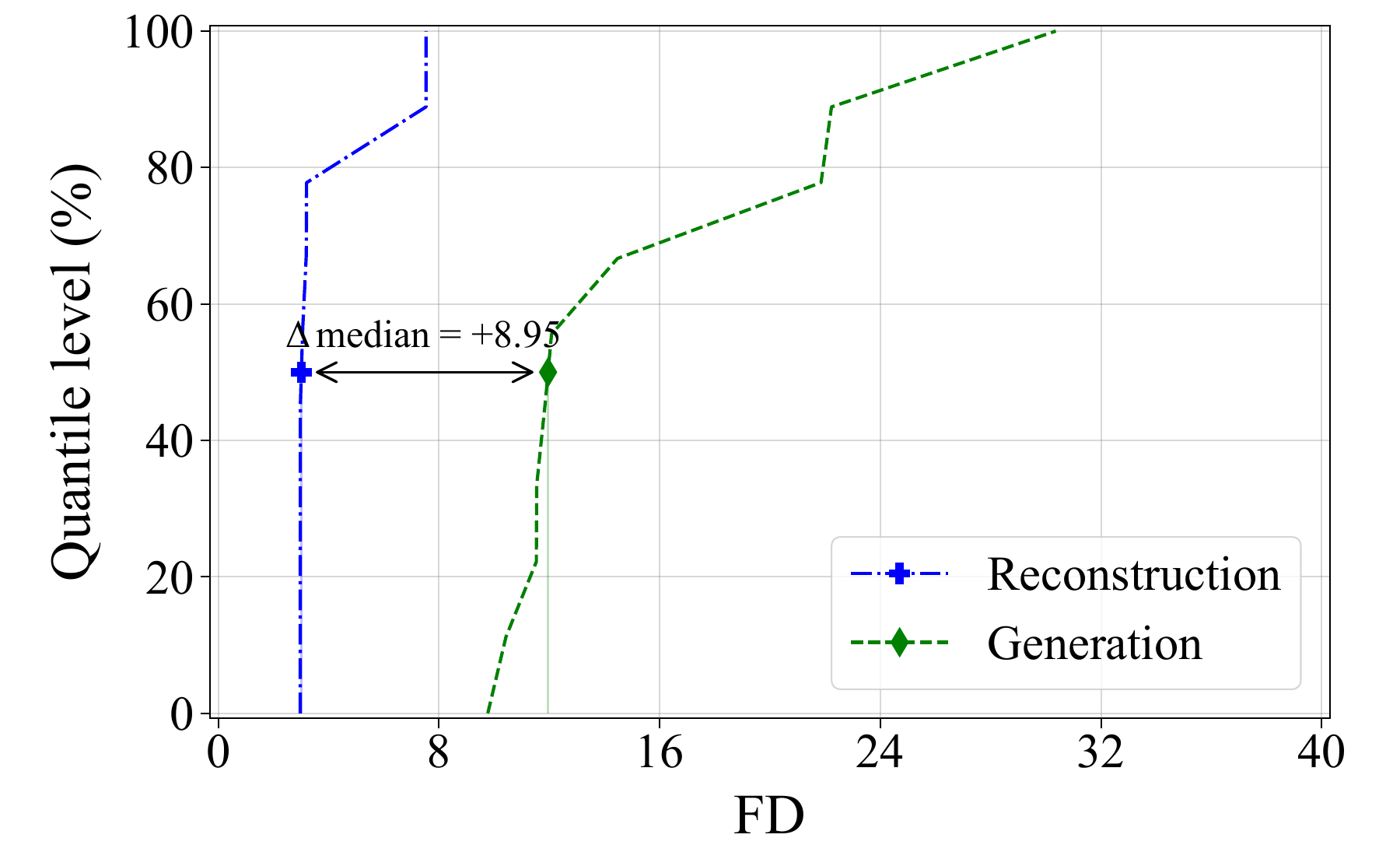}
\vspace{-9mm}
\caption{\small{Quantile comparison of rFD and gFD on AudioCaps. The two curves remain separated throughout.}}
\label{fig:fd-quantile-level-audiocaps}
\end{center}
\end{wrapfigure}

Figure~\ref{fig:fd-quantile-level-audiocaps} plots the empirical quantile curves of rFD and gFD across models on AudioCaps; the two curves remain separated throughout, with a median gap of 8.95. The same separation holds on MusicCaps and VGGSound-Omni (Figures~\ref{fig:fd-quantile-level-musiccaps}-\ref{fig:fd-quantile-level-vggsound-omni}, Appendix~\ref{app:additional-quantile}), where the median gaps are larger still, reaching 18.88 and 11.81 respectively. This consistent separation across the full quantile range on all three datasets shows that the reconstruction-generation gap is a broad phenomenon across the evaluated models, rather than being driven by a few individual cases.

Reconstruction and generation are evaluated on the same datasets and under the same evaluation protocol, and both pathways use the same codec decoder. However, generation additionally depends on the latent generative model, so the observed reconstruction–generation gap does not by itself isolate the effect of the decoder input-distribution shift. Together with the clear separation between $\mathcal{P}_e$ and $\mathcal{P}_g$ observed in Section 2.2, these results motivate a direct empirical question: can adapting the decoder to generation-relevant latents improve end-to-end generation performance? We study this question in the next section.
\section{Generation-Aware Codec Decoder Adaptation}
\label{sec:generation-aware-adaptation}

Standard neural audio codecs are trained to decode encoder-induced latents back to the source audio, with reconstruction-based supervision~\citep{Zeghidour2021SoundStreamAE,Defossez2022HighFN,Kumar2023HighFidelityAC,Evans2025StableAO}. This supervision is naturally available for encoder-induced latents but not for generator-produced latents, and naive adaptation on either comes with a fundamental drawback.

Adapting on encoder-induced latents $z_e \sim \mathcal{P}_e$ preserves correspondence with the source audio, but continues to train the decoder on the same latent distribution seen during codec training and therefore does not directly address the decoder train-generation mismatch. Adapting on generator-produced latents $z_g \sim \mathcal{P}_g$, in contrast, exposes the decoder to its generation-time inputs, but without a paired source waveform, reconstruction supervision is unavailable.

To resolve this, we introduce \textbf{AudioGAR}, which constructs intermediate latents between reconstruction and generation. These latents become progressively more generation-relevant while retaining correspondence with the source audio at lower noise levels, making them suitable for decoder adaptation. We first describe how AudioGAR latents are constructed, then show how they bridge $\mathcal{P}_e$ and $\mathcal{P}_g$, and finally present generation-aware codec decoder adaptation.

\subsection{AudioGAR Latent Construction}
\label{sec:audiogar-construction}

\begin{wrapfigure}[12]{r}{0.55\textwidth}
\small
\vspace{-5mm}
\begin{center}
\includegraphics[width=0.55\textwidth]{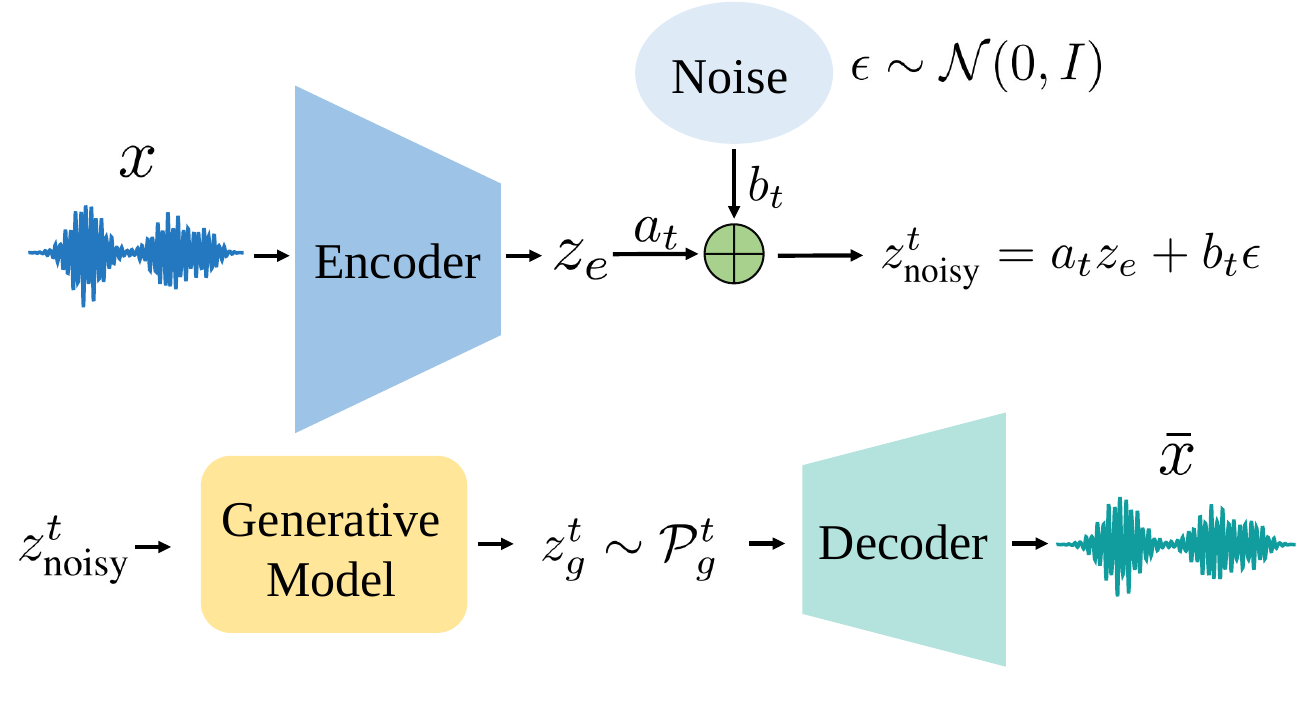}
\vspace{-7mm}
\caption{\small{Illustration of the AudioGAR process.}}
\label{fig:audio-gar-pipeline}
\end{center}
\end{wrapfigure}
AudioGAR constructs intermediate latents by perturbing an encoder-induced latent and then denoising it with the frozen latent diffusion model, as illustrated in Figure~\ref{fig:audio-gar-pipeline}. Given a source waveform $x$ and its encoder-induced latent $z_e=\mathcal{E}_{\theta}(x)$, we first add Gaussian noise according to the diffusion schedule, yielding $z_{\mathrm{noisy}}^{t}=a_t z_e+b_t\epsilon$, where $\epsilon\sim\mathcal{N}(0,I)$ and $t\in[0,1]$ denotes the normalized diffusion time. We characterize the perturbation strength by the relative noise level $\eta_t=b_t/(a_t+b_t)$. For the cosine parameterization used in our experiments, $a_t=\cos(\pi t/2)$ and $b_t=\sin(\pi t/2)$, so $\eta_t$ increases monotonically from 0 to 1 as $t$ goes from 0 to 1.

Starting from $z_{\mathrm{noisy}}^{t}$, the frozen latent diffusion model $\mathcal{G}_{\varphi}$ denoises the latent from diffusion time $t$ toward the clean latent space using $\lfloor N_{\max}t\rceil$ denoising steps. We denote the resulting latent by $z_g^t=\mathcal{G}_{\varphi}(z_{\mathrm{noisy}}^{t},t,c) \sim \mathcal{P}_g^t$, where $c$ is the conditioning information associated with the source audio. The codec encoder and latent diffusion model remain frozen throughout this construction.

The noise level $\eta_t$ controls the position between reconstruction and generation. At $t=0$, we have $\eta_t=0$, so no perturbation or denoising is applied and $z_g^t=z_e$, recovering the reconstruction endpoint. At $t=1$, we have $\eta_t=1$, so the encoder latent is fully replaced by Gaussian noise and the full $N_{\max}$-step generative process is applied, recovering the generation endpoint. Intermediate values of $t$ therefore produce a family of latents between these two endpoints. Because these latents originate from a particular encoder latent $z_e$, lower-noise AudioGAR latents retain correspondence with the source waveform while incorporating information from the generative process.

\subsection{Bridging Reconstruction and Generation}
\label{sec:audiogar-trajectory}

We next examine whether these intermediate latents indeed trace a trajectory between the encoder-induced and generator-produced latent distributions. Using AudioX~\citep{Tian2026AudioXTurboAU} on AudioCaps~\citep{Kim2019AudioCapsGC} and MusicCaps~\citep{Agostinelli2023MusicLMGM}, we extract encoder-induced latents $z_e \sim \mathcal{P}_e$, generator-produced latents $z_g \sim \mathcal{P}_g$, and AudioGAR latents $z_g^t \sim \mathcal{P}_g^t$ across different noise levels, and compare each intermediate distribution $\mathcal{P}_g^t$ with both endpoints.

Directly characterizing the discrepancy between high-dimensional latent distributions is difficult, so we use Latent Fr{\'e}chet Distance (Latent-FD) as a tractable approximation. Latent-FD computes the Fr{\'e}chet distance from the empirical means and covariances of two latent sets under a Gaussian approximation, following the same general principle as widely used Fr{\'e}chet metrics for image and audio generation. We provide the full definition in Appendix~\ref{app:latent-fd}.

Figures~\ref{fig:latent-fd-audiocaps} and~\ref{fig:latent-fd-musiccaps} show the resulting distances on AudioCaps and MusicCaps. As the noise level increases, the Latent-FD between $\mathcal{P}_g^t$ and $\mathcal{P}_e$ increases monotonically, while the Latent-FD between $\mathcal{P}_g^t$ and $\mathcal{P}_g$ decreases monotonically, consistently across both datasets. AudioGAR thus provides a smooth distributional trajectory from reconstruction toward generation, rather than an abrupt transition, supporting its use for generation-aware decoder adaptation.

\begin{figure*}[t]
    \vspace{-2mm}
	\centering
    \captionsetup[subfloat]{skip=-0.5pt}
	\subfloat[AudioCaps]{ 
		\label{fig:latent-fd-audiocaps}  
		\includegraphics[width=0.485\textwidth]{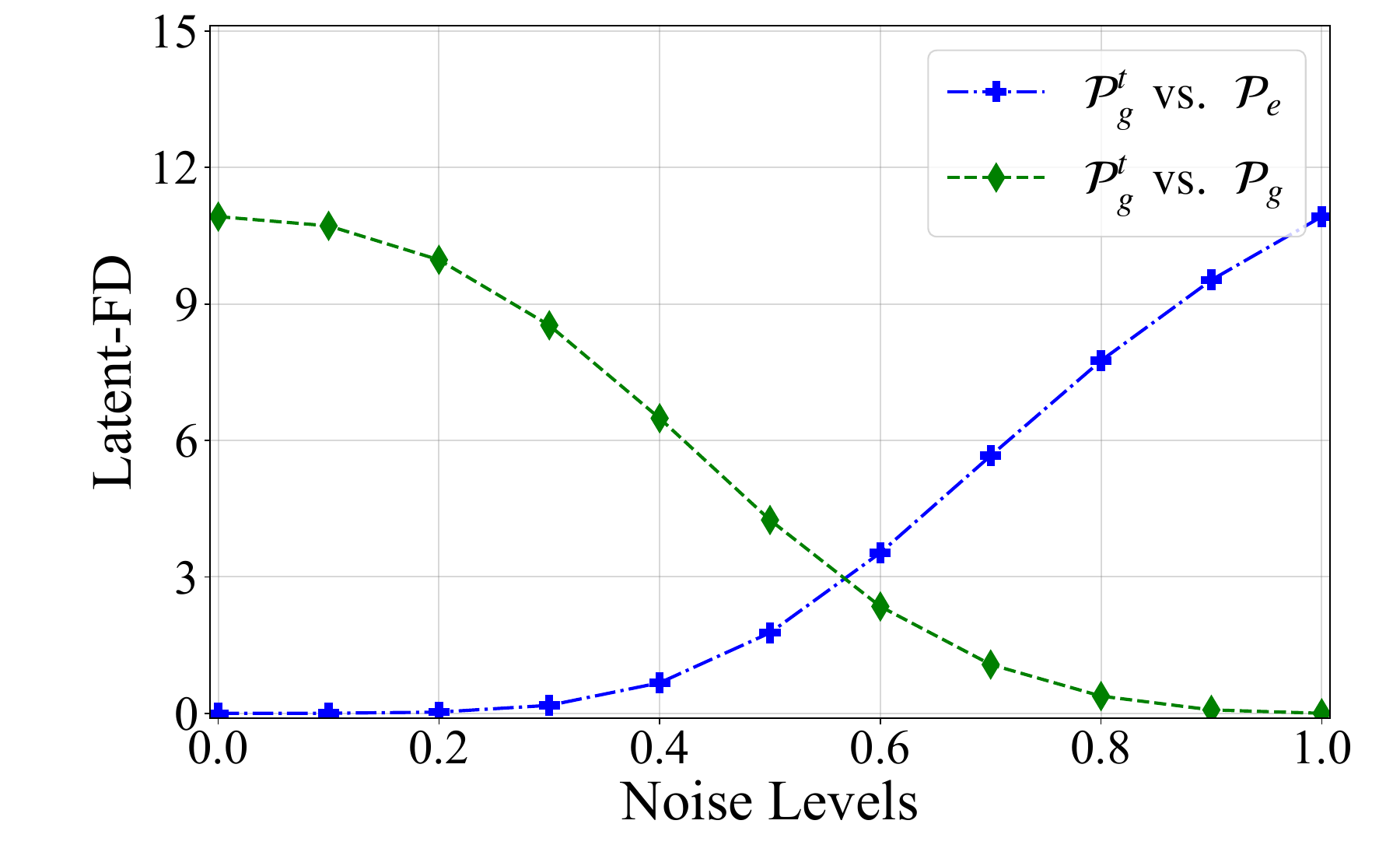}
    }
	\subfloat[MusicCaps]{
		\label{fig:latent-fd-musiccaps}
		\includegraphics[width=0.485\textwidth]{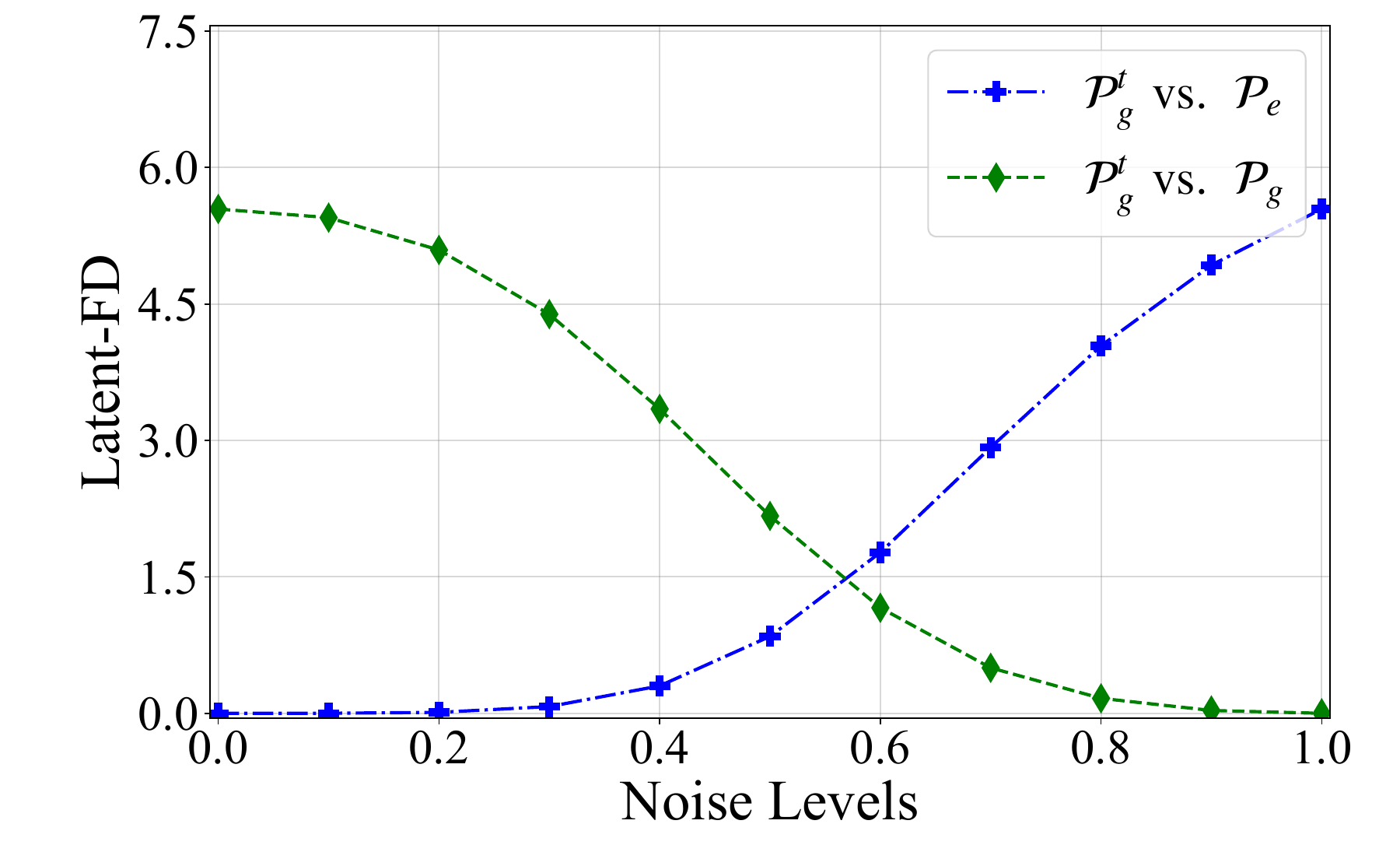}
	}
    \vspace{-2ex}
    \caption{\small{\textbf{AudioGAR bridges encoder-induced and generator-produced latent distributions.} Latent-FD between the intermediate AudioGAR distribution $\mathcal{P}_g^t$ and the two endpoints, $\mathcal{P}_e$ and $\mathcal{P}_g$, across noise levels on (a) AudioCaps and (b) MusicCaps. As the noise level increases, $\mathcal{P}_g^t$ moves farther from $\mathcal{P}_e$ and closer to $\mathcal{P}_g$, tracing a trajectory from reconstruction toward generation.}}
    \label{fig:latent-fd}
    \vspace{-3ex}
\end{figure*}

\subsection{Decoder Adaptation with AudioGAR Latents}
\label{sec:codec-adaptation}

We fine-tune the pretrained codec decoder on AudioGAR latents, keeping the same decoder-side training objective as the original codec. Throughout adaptation, the codec encoder $\mathcal{E}_{\theta}$ and latent generative model $\mathcal{G}_{\varphi}$ are frozen, while the pretrained decoder $\mathcal{D}_{\phi}$ is updated. The key difference from standard codec training is the decoder input: instead of encoder-induced latents $z_e$, the decoder receives AudioGAR latents $z_g^t$, which retain correspondence with the source audio while shifting toward the generator-produced latent distribution.

For a source waveform $x$ with AudioGAR latent $z_g^t$, the reconstructed waveform $\bar{x}=\mathcal{D}_{\phi}(z_g^t)$ is optimized under $\mathcal{L}_{\text{AudioGAR}}=\mathcal{L}_{\text{codec}}(x,\bar{x})$, where $\mathcal{L}_{\text{codec}}$ denotes the decoder-side reconstruction and adversarial objectives of the corresponding pretrained codec. For EnCodec-style codec training, these include time- and frequency-domain reconstruction losses, adversarial losses, and feature-matching losses~\citep{Defossez2022HighFN}; for the AudioLDM VAE~\citep{Liu2023AudioLDMTG}, mel-spectrogram reconstruction and PatchGAN-based adversarial supervision~\citep{Isola2017ImagetoImageTW}. All loss weighting hyperparameters are kept identical to those used in the original codec training.  We refer to this decoder adaptation procedure using AudioGAR latents as \textbf{AudioGAR}.

AudioGAR relies on a generative process that can start from a partially perturbed encoder latent and progressively denoise it toward the generated latent distribution. It is therefore designed for latent diffusion and flow-matching models~\citep{Rombach2022HighResolutionIS,Peebles2022ScalableDM,Geng2025ImprovedMF,Liu2023AudioLDMTG,Ghosal2023TexttoAudioGU,Tian2026AudioXTurboAU}, and does not directly apply to autoregressive latent generators such as AudioGen~\citep{Kreuk2023AudioGenTG} and MusicGen~\citep{Copet2023SimpleAC}, or to other generators without an analogous intermediate denoising trajectory.

\section{Experiments}
\label{sec:experiments}
We evaluate AudioGAR on pretrained latent audio generative models to study its effectiveness and behavior during codec decoder adaptation. We first describe the experimental setup and then report the main results. Finally, we present sensitivity analysis, ablation studies, and training dynamics. Code and checkpoints will be publicly released to ensure reproducibility.

\subsection{Experimental Setup}
\label{sec:experimemtal-setup}

\vspace{-2pt}
\paragraph{Models and Training Setup.}
We primarily evaluate AudioGAR on AudioX~\citep{Tian2026AudioXTurboAU} and additionally apply it to TangoMusic~\citep{Kong2024ImprovingTM}. We initialize from released pretrained checkpoints and freeze the codec encoder and latent generative model in both cases. For AudioX, we fine-tune only the codec decoder; for TangoMusic, we jointly fine-tune the VAE decoder and HiFi-GAN vocoder. AudioGAR latents $z_g^t$ are constructed with $\eta_t=0.2$. Since the released checkpoints do not include the discriminators used for codec training, we initialize the discriminator with random weights and train them jointly with the decoder-side components following the respective codec training procedures~\citep{Evans2025StableAO,Kong2020HiFiGAN}. AudioX adaptation runs for 100k steps using 508.3 hours of audio. As shown in Table~\ref{tab:audiox_audiogar_efficiency}, this amounts to 1.5\% of the original training audio hours and 0.26\% of the original training cost. Full training details are provided in Appendix~\ref{app:experimental-details}.

\vspace{-2pt}
\paragraph{Evaluation Metrics.} Following AudioX~\citep{Tian2026AudioXTurboAU}, we evaluate overall audio generation performance using several established metrics: Kullback-Leibler Divergence (KL), Inception Score (IS), Fréchet Distance (FD) computed with PANNs embeddings~\citep{Kong2019PANNsLP}, Fréchet Audio Distance (FAD) computed with VGGish embeddings~\citep{Hershey2016CNNAF}, and Production Complexity (PC) and Production Quality (PQ)~\citep{Tjandra2025MetaAA}. Additional details on these metrics are provided in Appendix~\ref{app:metrics}.

\vspace{-6pt}
\paragraph{Baselines.}
For text-to-audio task on AudioCaps~\citep{Kim2019AudioCapsGC}, we compare against AudioGen~\citep{Kreuk2023AudioGenTG}, AudioLDM-L-Full and AudioLDM-S-Full~\citep{Liu2023AudioLDMTG}, Tango 2~\citep{Majumder2024Tango2A}, Tango (AF-AC, FT AudioCaps)~\citep{Kong2024ImprovingTM}, MAGNET-large~\citep{Ziv2024MaskedAG}, MMAudio~\citep{Cheng2024MMAudioTM}, AudioLDM-2-Base and AudioLDM-2-Large~\citep{Liu2024AudioLDM2L}, Stable Audio Open~\citep{Evans2025StableAO}, and AudioX~\citep{Tian2026AudioXTurboAU}. For the text-to-music task on MusicCaps~\citep{Agostinelli2023MusicLMGM}, we compare against MusicGen~\citep{Copet2023SimpleAC}, AudioLDM-L-Full~\citep{Liu2023AudioLDMTG}, MAGNET-large~\citep{Ziv2024MaskedAG}, AudioLDM-2-Base and AudioLDM-2-Large~\citep{Liu2024AudioLDM2L}, Stable Audio Open\\ ~\citep{Evans2025StableAO}, Tango and Tango Full (FT AudioCaps + MusicCaps)~\citep{Ghosal2023TexttoAudioGU}, TangoMusic~\citep{Kong2024ImprovingTM}, and AudioX~\citep{Tian2026AudioXTurboAU}.

\subsection{Main Results}
\label{sec:main-results}
\paragraph{AudioGAR on AudioX.}
Table~\ref{tab:main_table} reports the main results on AudioCaps and MusicCaps. Compared with our reproduced AudioX baseline, AudioX + AudioGAR improves both FD and FAD on the two datasets. On AudioCaps, FD decreases from 11.83 to 11.62 and FAD from 1.61 to 1.34, corresponding to relative reductions of 1.8\% and 16.8\%, respectively. On MusicCaps, the improvements are larger: FD drops from 9.56 to 8.33 and FAD from 1.60 to 1.10, reductions of 12.9\% and 31.3\%. Performance on KL, IS, PC, and PQ remains broadly matched with the pretrained AudioX baseline. Among the methods in Table~\ref{tab:main_table}, AudioX + AudioGAR achieves the best FAD on AudioCaps and the best FD and FAD on MusicCaps, with only marginal changes in the other evaluated metrics. Notably, decoder adaptation introduces no additional inference cost, as only the decoder weights are updated while the inference pipeline and generative model remain unchanged.

\paragraph{Generalization to TangoMusic.}
We further apply AudioGAR to TangoMusic~\citep{Kong2024ImprovingTM} to test whether the same adaptation strategy transfers to a different latent audio generative model and codec pipeline. Relative to our reproduced TangoMusic baseline, TangoMusic + AudioGAR reduces FD from 15.20 to 14.25 and FAD from 1.85 to 1.47 on MusicCaps, improvements of 6.3\% and 20.5\%, respectively. The remaining metrics stay closely matched, with KL unchanged and only minor differences in IS, PC, and PQ. These results confirm that the gains from AudioGAR are not specific to AudioX, and that generation-aware decoder adaptation transfers across different pretrained latent audio generative models and decoder architectures.

\begin{table*}[t]
\centering
\small
\caption{\small{\textbf{Performance evaluation across tasks and datasets.} 
T2A and T2M denote Text-to-Audio and Text-to-Music, respectively. 
$^{\dagger}$ denotes results reported in AudioX~\citep{Tian2026AudioXTurboAU}, while $^{\ddagger}$ denotes results from our experiments, including reproduced baselines and AudioGAR evaluations. The best result for each metric within each task is shown in bold.}}
\vspace{-2.5ex}
\resizebox{0.95\textwidth}{!}{%
\begin{tabular}{l|l|l|cccccc}
\toprule
Dataset & Method & Task & KL ({\color{green!60!black}\bfseries$\pmb{\downarrow}$}) & IS ({\color{green!60!black}\bfseries$\pmb{\uparrow}$}) & FD ({\color{green!60!black}\bfseries$\pmb{\downarrow}$}) & FAD ({\color{green!60!black}\bfseries$\pmb{\downarrow}$}) & PC ({\color{green!60!black}\bfseries$\pmb{\uparrow}$}) & PQ ({\color{green!60!black}\bfseries$\pmb{\uparrow}$}) \\
\midrule
\multirow{15}{*}{\shortstack[c]{\centering AudioCaps}}
& AudioGen$^{\dagger}$  & T2A  & 1.39 & 10.22 & 13.29 & 1.72 & 3.26 & 5.25  \\
& AudioLDM-L-Full$^{\dagger}$ & T2A  & 2.00 & 6.51 & 37.27 & 8.37 & 2.82 & 5.67 \\
& AudioLDM-S-Full$^{\ddagger}$ & T2A  & 1.52 & 7.59 & 21.85 & 4.79 & 3.21 & 5.81\\
& Tango 2$^{\dagger}$ & T2A  & \textbf{1.11} & 10.37 & 12.22 & 3.20 & \textbf{3.63} & 5.82 \\
& Tango (AF-AC, FT AudioCaps)$^{\ddagger}$  & T2A & 1.24 & 10.66 & 12.08 & 2.55 & 3.44 & 6.13 \\ 
& MAGNET-large$^{\dagger}$ & T2A  & 1.62 & 7.46 & 24.88 & 2.99 & 3.25 & 5.15 \\
& MMAudio$^{\dagger}$ & T2A  & 1.35 & 12.03 & 12.63 & 4.71 & 3.06 & 5.64 \\
& AudioLDM-2-Base$^{\ddagger}$ & T2A  & 1.28 & 10.56 & 11.54 & 2.04 & 3.12 & 5.89 \\
& AudioLDM-2-Large$^{\dagger}$ & T2A  & 1.49 & 8.46 & 26.34 & 1.97 & 2.86 & 5.77 \\
& AudioLDM-2-Large$^{\ddagger}$ & T2A  & 1.22 & 10.15 & 11.53 & 1.86 & 3.14 & 5.80 \\
& Stable Audio Open$^{\dagger}$ & T2A  & 2.01 & 10.37 & 29.01 & 3.15 & 2.77 & \textbf{6.16} \\
& Stable Audio Open$^{\ddagger}$ & T2A  & 2.04 & 10.12 & 30.36 & 3.41 & 2.69 & 6.08 \\
& AudioX$^{\dagger}$ & T2A  & 1.27 & \textbf{12.48} & \textbf{11.51} & 1.59 & 3.32 & 5.80 \\
& AudioX$^{\ddagger}$ & T2A  & 1.31 & 12.47 & 11.83 & 1.61 & 3.16 & 5.73 \\
&\cellcolor{cyan!20}\textbf{AudioX + AudioGAR (Ours)}$^{\ddagger}$ &\cellcolor{cyan!20}T2A  &\cellcolor{cyan!20}1.29 &\cellcolor{cyan!20}12.06 &\cellcolor{cyan!20}11.62 &\cellcolor{cyan!20}\textbf{1.34} &\cellcolor{cyan!20}3.11 &\cellcolor{cyan!20}5.73 \\
\midrule    
\multirow{16}{*}{\shortstack[c]{\centering MusicCaps}}
& MusicGen$^{\dagger}$       & T2M  & 1.43 & 2.24 & 25.40 & 4.55 & 5.19 & 7.16 \\
& AudioLDM-L-Full$^{\dagger}$ & T2M  & 1.45 & 2.49 & 34.44 & 6.34 & 4.72 & 6.10 \\
& MAGNET-large$^{\dagger}$ & T2M  & 1.32 & 1.98 & 23.88 & 4.24 & 5.84 & 6.71\\
& AudioLDM-2-Base$^{\ddagger}$ & T2M & 1.20 & 2.51 & 21.71 & 3.82 & 5.35 & \textbf{7.39} \\
& AudioLDM-2-Large$^{\dagger}$ & T2M  & 1.26 & 2.84 & 15.61 & 2.80 & 5.22 & 6.70 \\
& AudioLDM-2-Large$^{\ddagger}$ & T2M  & 1.17 & 2.70 & 18.84 & 2.51 & 5.38 & 7.23 \\
& Stable Audio Open$^{\dagger}$  & T2M  & 1.51 & 2.94 & 36.33 & 3.23 & 3.91 & 7.18 \\
& Stable Audio Open$^{\ddagger}$ & T2M  & 1.56 & 2.98 & 37.72 & 3.42 & 3.75 & 7.26 \\
& Tango$^{\ddagger}$ & T2M  & 2.03 & 2.05 & 46.80 & 5.64 & \textbf{5.86} & 6.64 \\
& Tango Full (FT AudioCaps + MusicCaps)$^{\ddagger}$ & T2M  & 0.98 & 2.71 & 20.43 & 3.61 & 5.71 & 6.29 \\
& TangoMusic$^{\dagger}$ & T2M  & 1.13 & 2.86 & 15.00 & 1.88 & 5.57 & 7.06 \\
& TangoMusic$^{\ddagger}$ & T2M  & 1.09 & 2.85 & 15.20 & 1.85 & 5.62 & 7.22 \\
&\cellcolor{cyan!20}\textbf{TangoMusic + AudioGAR (Ours)}$^{\ddagger}$ &\cellcolor{cyan!20}T2M  &\cellcolor{cyan!20}1.09 &\cellcolor{cyan!20}2.83 &\cellcolor{cyan!20}14.25 &\cellcolor{cyan!20}1.47 &\cellcolor{cyan!20}5.61 &\cellcolor{cyan!20}7.16 \\
& AudioX$^{\dagger}$  & T2M  & \textbf{0.96} &  3.55 & 9.76 & 1.53 & 5.21 & 6.70 \\
& AudioX$^{\ddagger}$  & T2M  & 1.00 & \textbf{3.65} & 9.56 & 1.60 & 4.78 & 6.61 \\
&\cellcolor{cyan!20}\textbf{AudioX + AudioGAR (Ours)}$^{\ddagger}$  &\cellcolor{cyan!20}T2M  &\cellcolor{cyan!20}0.99 & \cellcolor{cyan!20}3.59 &\cellcolor{cyan!20}\textbf{8.33} &\cellcolor{cyan!20}\textbf{1.10} &\cellcolor{cyan!20}4.75 &\cellcolor{cyan!20}6.56 \\
\midrule
\end{tabular}
}
\vspace{-2ex}
\label{tab:main_table}
\end{table*}

\begin{table*}[t]
\centering
\small
\caption{\small{\textbf{Effect of latent choice during decoder adaptation.}
We compare the pretrained AudioX baseline with decoder adaptation using either encoder-induced latents $z_e$ or AudioGAR latents $z_g^t$ on AudioCaps and MusicCaps. \XSolidBrush and \CheckmarkBold indicate whether decoder adaptation is applied.}}
\vspace{-2.5ex}
\resizebox{0.95\textwidth}{!}{%
\begin{tabular}{l|cc|cccccc}
\toprule
Dataset & Decoder Adaptation & Decoder Latents & KL ({\color{green!60!black}\bfseries$\pmb{\downarrow}$}) & IS ({\color{green!60!black}\bfseries$\pmb{\uparrow}$}) & FD ({\color{green!60!black}\bfseries$\pmb{\downarrow}$}) & FAD ({\color{green!60!black}\bfseries$\pmb{\downarrow}$}) & PC ({\color{green!60!black}\bfseries$\pmb{\uparrow}$}) & PQ ({\color{green!60!black}\bfseries$\pmb{\uparrow}$}) \\
\midrule
\multirow{3}{*}{\shortstack[c]{\centering AudioCaps}}
& \XSolidBrush & \Large -  & 1.31 & \textbf{12.47} & 11.83 & 1.61 & \textbf{3.16} & \textbf{5.73} \\
& \CheckmarkBold & \Large $z_e$ & \textbf{1.29} & 12.19 & 11.72 & 1.55 & 3.14 & 5.70\\
& \CheckmarkBold & \Large $z_g^t$ & \textbf{1.29} & 12.06 & \textbf{11.62} & \textbf{1.34} & 3.11 & \textbf{5.73} \\
\midrule    
\multirow{3}{*}{\shortstack[c]{\centering MusicCaps}}
& \XSolidBrush & \Large -  & 1.00 & \textbf{3.65} & 9.56 & 1.60 & \textbf{4.78} & \textbf{6.61} \\
& \CheckmarkBold & \Large $z_e$  & \textbf{0.99} & 3.62 & 8.87 & 1.38  &  4.76 & 6.50 \\
& \CheckmarkBold & \Large $z_g^t$ &  \textbf{0.99} & 3.59 & \textbf{8.33} & \textbf{1.10} &4.75 & 6.56 \\
\midrule
\end{tabular}
}
\label{tab:ablation_table}
\end{table*}

\begin{figure*}[!t]
    \vspace{-2mm}
	\centering
    \captionsetup[subfloat]{skip=-0.5pt}
	\subfloat[AudioCaps]{ 
		\label{fig:sensitivity-fd-audiocaps}  
		\includegraphics[width=0.485\textwidth]{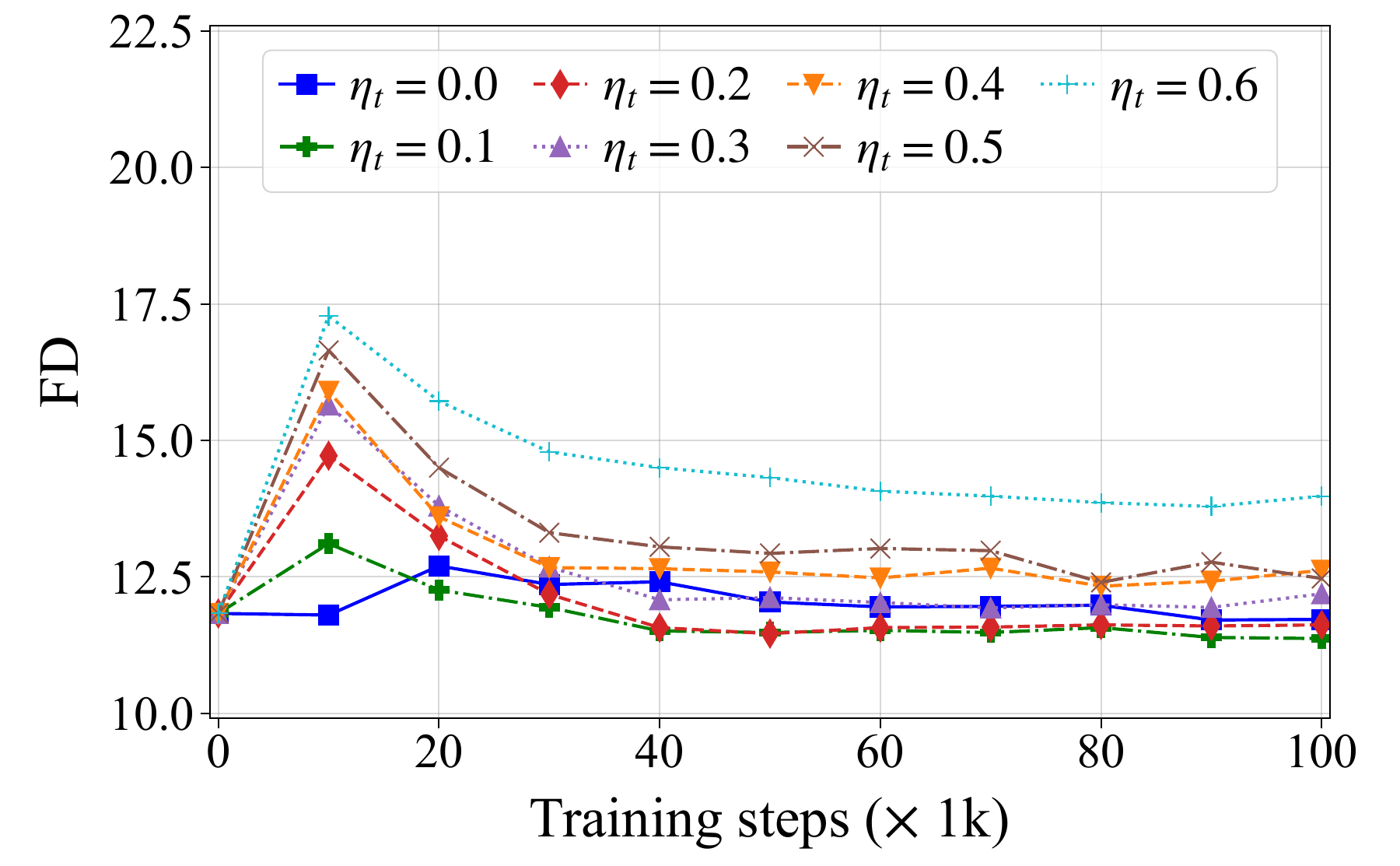}
    }
	\subfloat[MusicCaps]{
		\label{fig:sensitivity-fd-musiccaps}
		\includegraphics[width=0.485\textwidth]{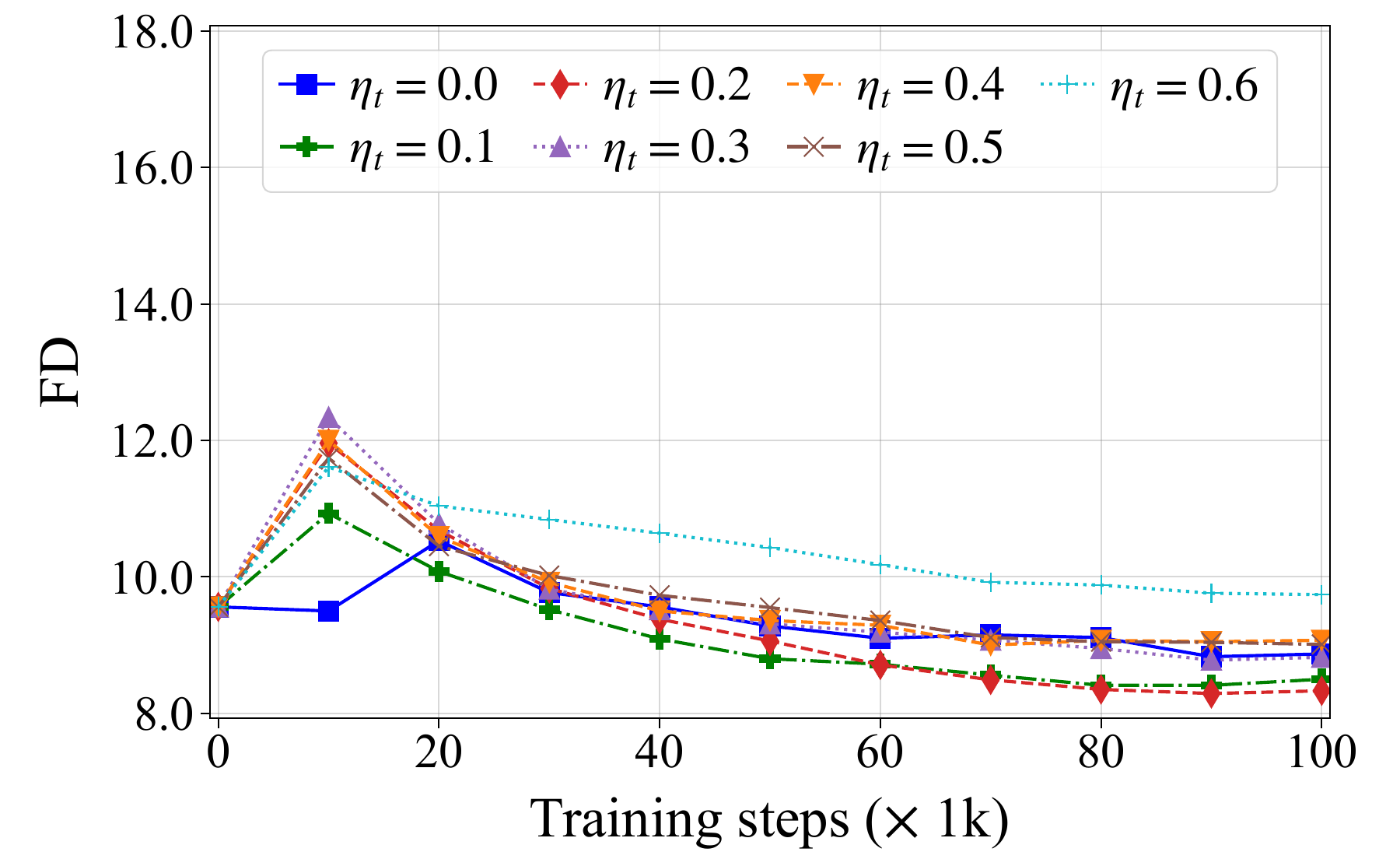}
	}
    \vspace{-2ex}
    \caption{\small{\textbf{Sensitivity of FD to the AudioGAR noise level.}
    FD throughout decoder adaptation with AudioX for different noise levels $\eta_t$ on (a) AudioCaps and (b) MusicCaps.}}
    \label{fig:sensitivity-fd}
\end{figure*}

\begin{figure*}[!t]
    \vspace{-2mm}
	\centering
    \captionsetup[subfloat]{skip=-0.5pt}
	\subfloat[AudioCaps]{ 
		\label{fig:sensitivity-fad-audiocaps}  
		\includegraphics[width=0.485\textwidth]{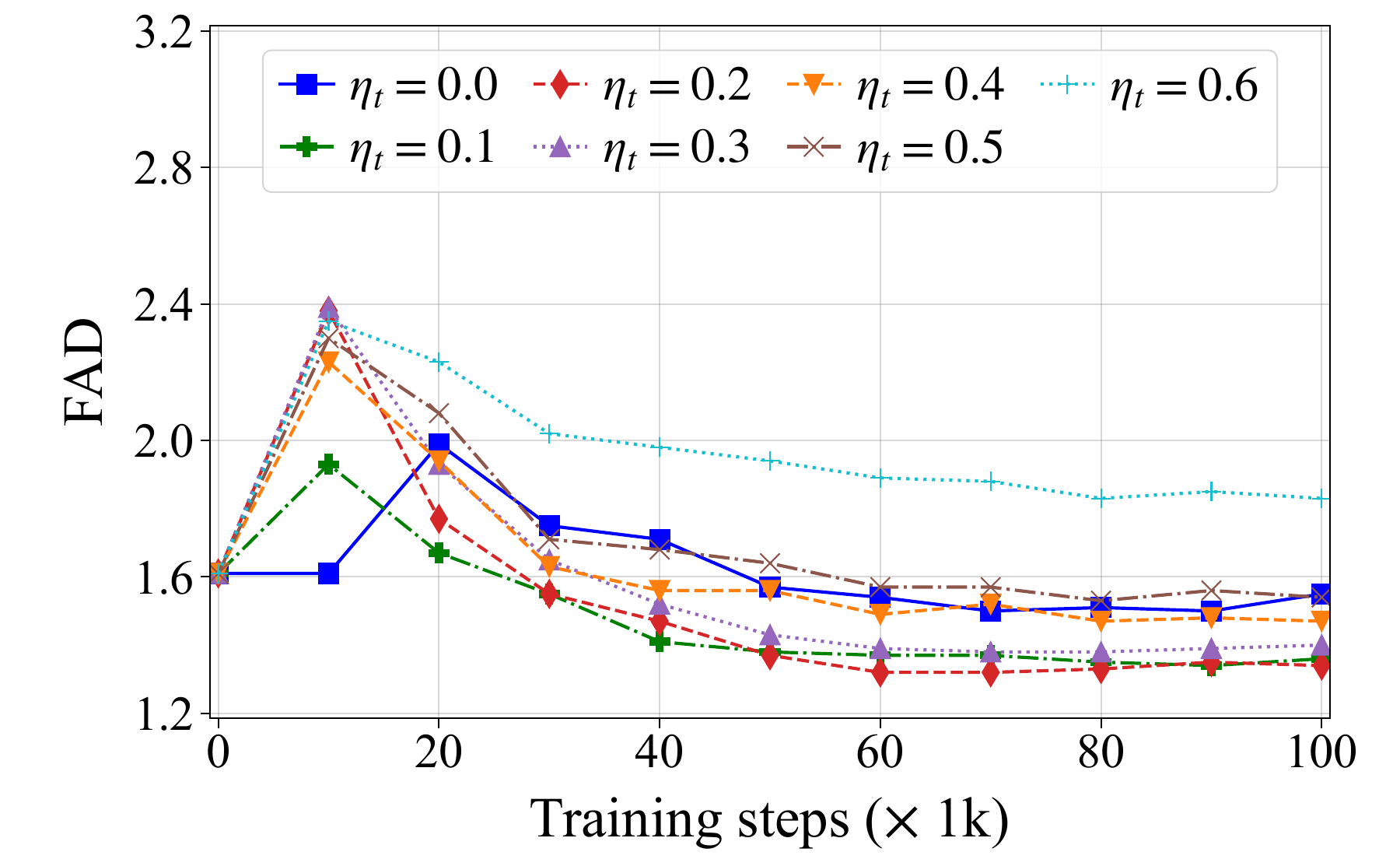}
    }
	\subfloat[MusicCaps]{
		\label{fig:sensitivity-fad-musiccaps}
		\includegraphics[width=0.485\textwidth]{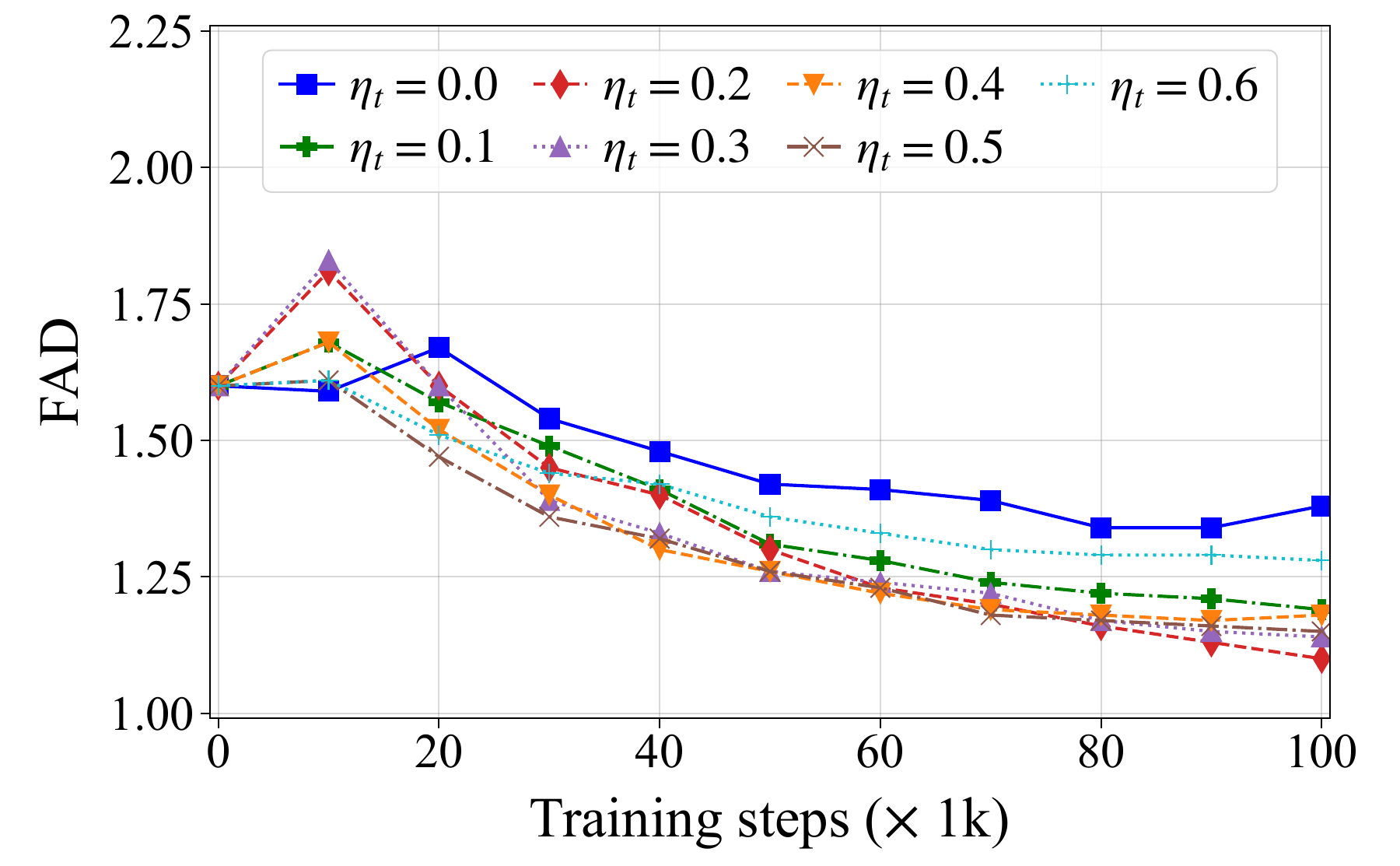}
	}
    \vspace{-2ex}
    \caption{\small{\textbf{Sensitivity of FAD to the AudioGAR noise level.}
    FAD throughout decoder adaptation with AudioX for different noise levels $\eta_t$ on (a) AudioCaps and (b) MusicCaps.}}
    \label{fig:sensitivity-fad}
    \vspace{-1ex}
  \end{figure*}

\subsection{Ablation and Analysis}
\label{sec:analysis}
We further analyze AudioGAR from three perspectives: the role of AudioGAR latents in decoder adaptation, sensitivity to the noise level, and the training dynamics of generation performance.

\paragraph{Effect of AudioGAR Latents.}
We first isolate the effect of using AudioGAR latents during decoder adaptation. Table~\ref{tab:ablation_table} compares the pretrained AudioX baseline with decoder adaptation using either encoder-induced latents $z_e$ or AudioGAR latents $z_g^t$, while keeping the adaptation procedure otherwise unchanged. Adapting on $z_e$ already provides modest improvements in FD and FAD, indicating that decoder fine-tuning itself can be beneficial. Replacing $z_e$ with $z_g^t$, however, yields consistently larger gains on both datasets. On AudioCaps, FD/FAD improve from 11.72/1.55 with $z_e$ to 11.62/1.34 with $z_g^t$; on MusicCaps, they improve from 8.87/1.38 to 8.33/1.10. This comparison isolates the benefit of exposing the decoder to generation-aware AudioGAR latents rather than simply continuing adaptation on the encoder-induced latent distribution.

\paragraph{Sensitivity to the Noise Level.}
We next study how the AudioGAR noise level affects decoder adaptation. Figures~\ref{fig:sensitivity-fd} and~\ref{fig:sensitivity-fad} show FD and FAD throughout 100k adaptation steps for $\eta_t\in\{0.0,0.1,\ldots,0.6\}$ on AudioCaps and MusicCaps. Intermediate noise levels, particularly $\eta_t=0.2$ and $\eta_t=0.3$, generally achieve better generation performance than either endpoint of the evaluated range. When $\eta_t$ is too small, such as $\eta_t=0.0$, the decoder continues to receive latents close to the encoder-induced distribution and therefore does not sufficiently address the decoder train-generation mismatch. Increasing $\eta_t$ makes the latents more generation-relevant, but excessively large noise levels, such as $\eta_t=0.6$, weaken their correspondence with the source audio and make paired reconstruction supervision less informative. These results highlight the trade-off underlying AudioGAR: intermediate noise levels provide generation-relevant latents while retaining sufficient source correspondence for decoder adaptation.

\begin{figure*}[!t]
    \vspace{-2mm}
	\centering
    \captionsetup[subfloat]{skip=-0.5pt}
	\subfloat[FD on AudioCaps]{ 
		\label{fig:dynamics-fd-audiocaps}  
		\includegraphics[width=0.225\textwidth]{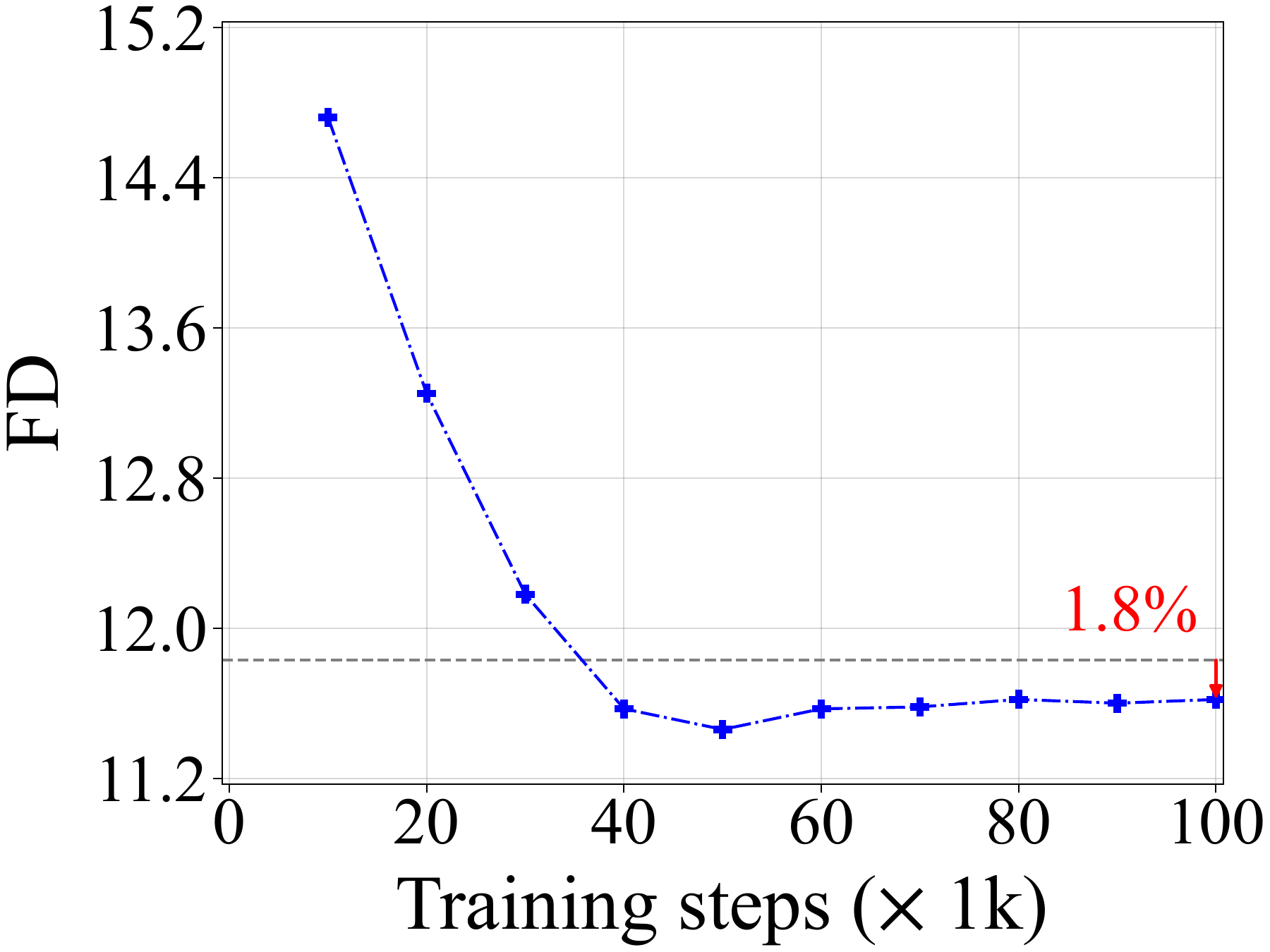}
    }
	\subfloat[FD on MusicCaps]{
		\label{fig:dynamics-fd-musiccaps}
		\includegraphics[width=0.225\textwidth]{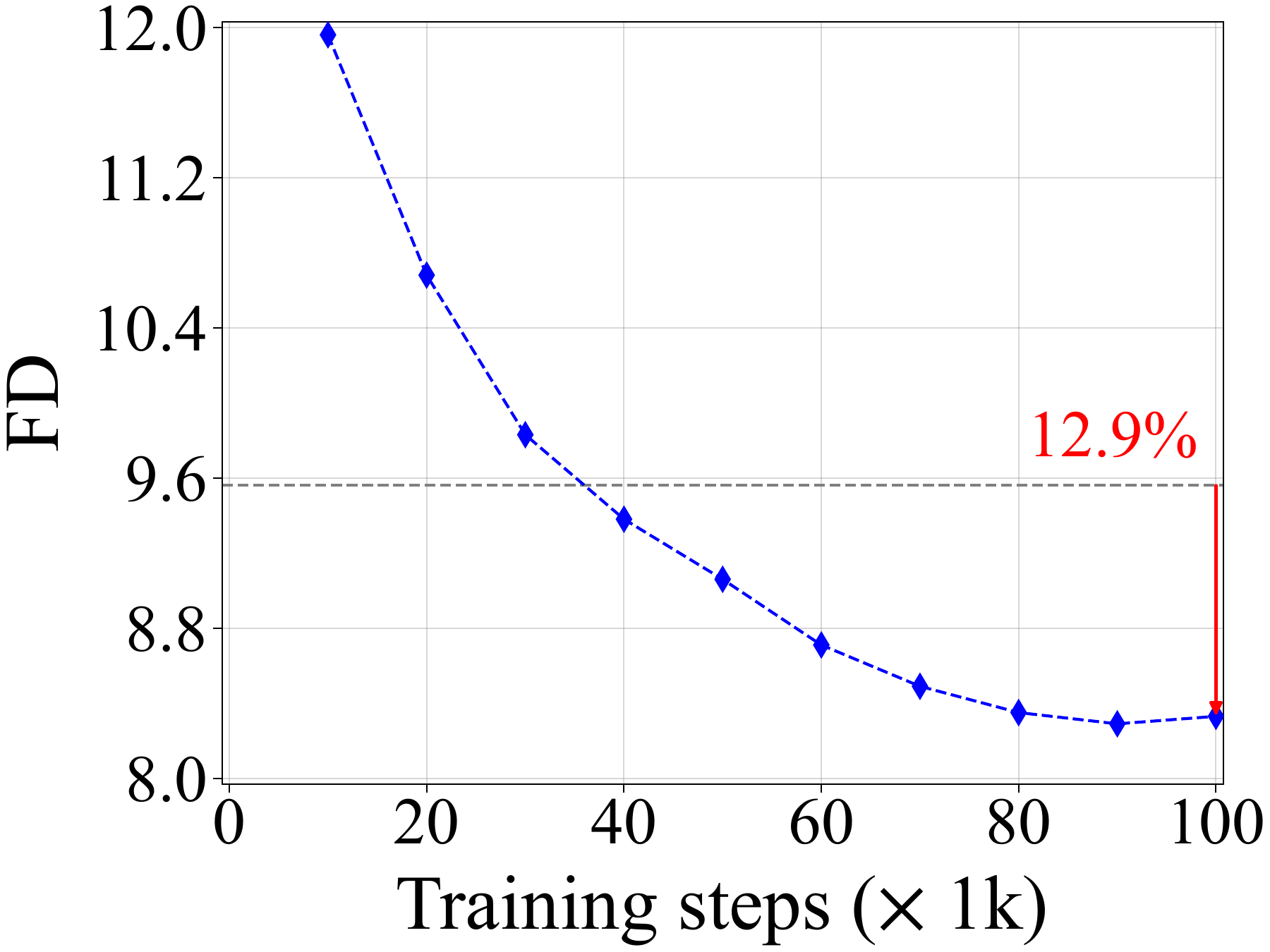}
	}
    \subfloat[FAD on AudioCaps]{ 
		\label{fig:dynamics-fad-audiocaps}  
		\includegraphics[width=0.225\textwidth]{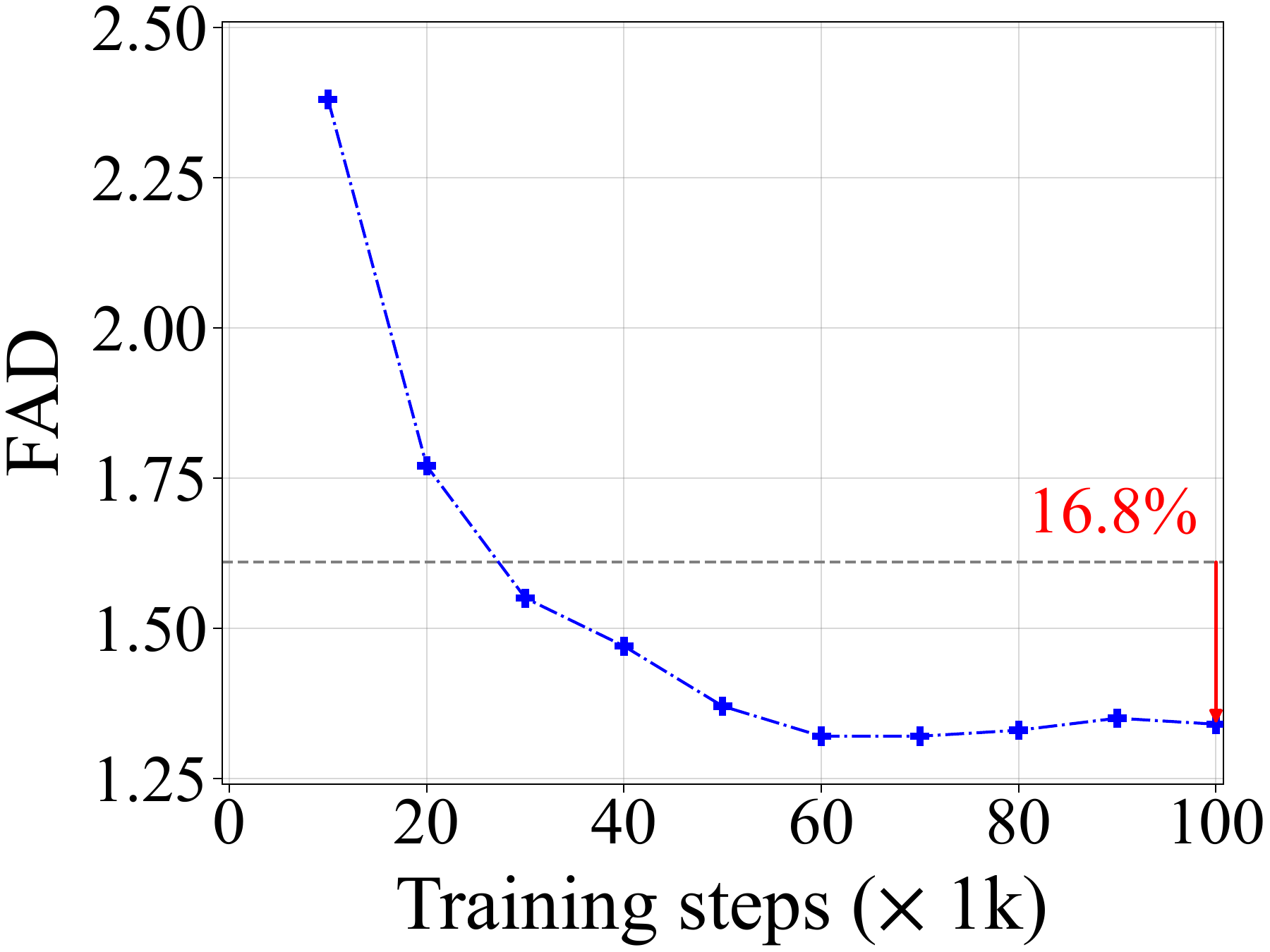}
    }
	\subfloat[FAD on MusicCaps]{
		\label{fig:dynamics-fad-musiccaps}
		\includegraphics[width=0.225\textwidth]{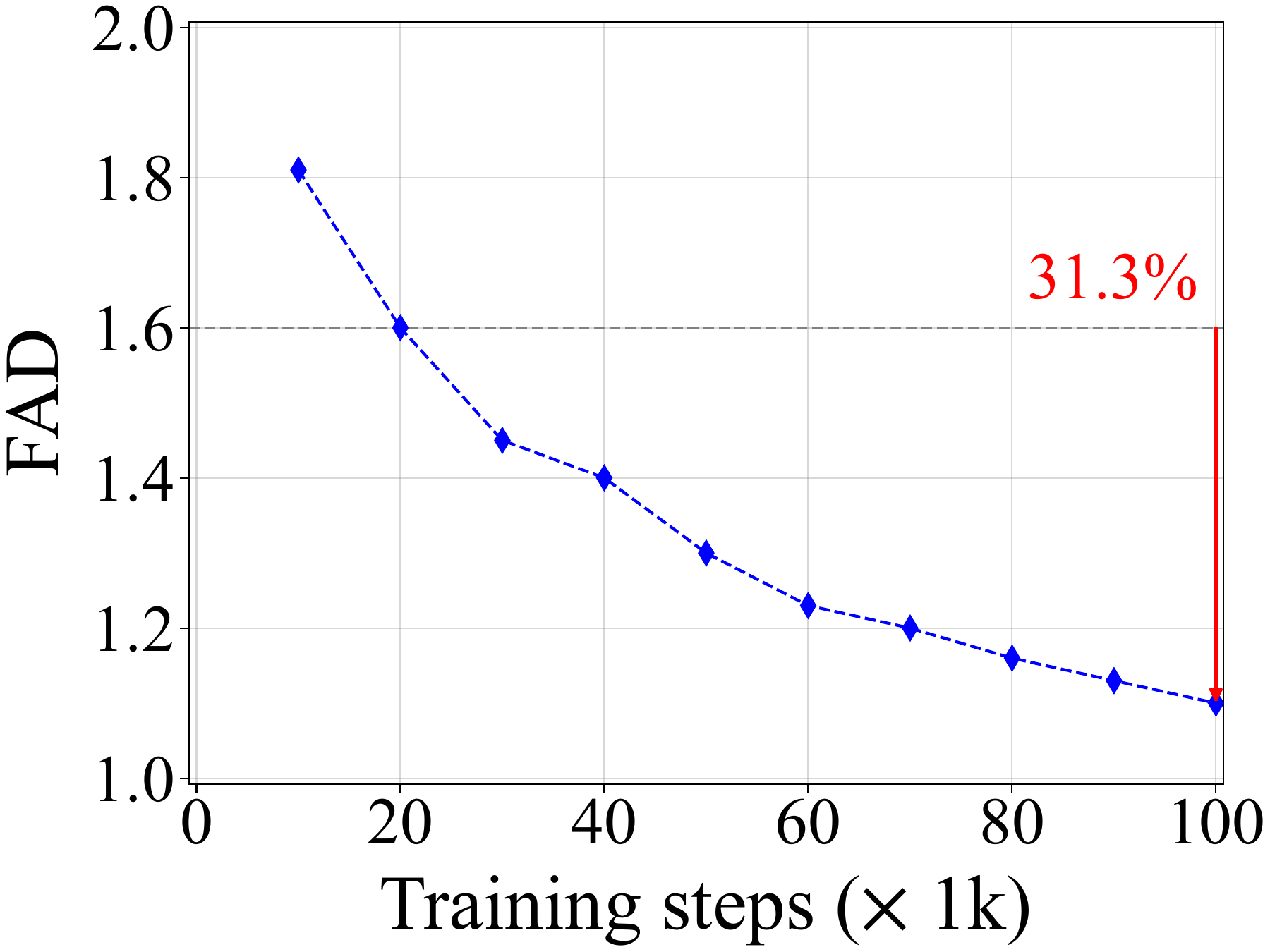}
	}
    \vspace{-1ex}
    \caption{\small{\textbf{Training dynamics of AudioGAR on AudioX.}
    FD and FAD over 100k decoder-adaptation steps on AudioCaps and MusicCaps with $\eta_t=0.2$.}}
    \label{fig:training-dynamics}
    \vspace{-3ex}
\end{figure*}

\paragraph{Training Dynamics.}
Finally, we examine how generation performance evolves during AudioGAR adaptation. Figure~\ref{fig:training-dynamics} tracks FD and FAD on AudioCaps and MusicCaps from the pretrained AudioX checkpoint at 0k steps through 100k adaptation steps. All four metrics initially worsen before improving with continued training and eventually surpassing the pretrained baseline. A possible explanation for the early decline is the random initialization of the discriminator, which is trained jointly with the pretrained decoder and may require time to provide useful adversarial supervision. The curves therefore complement the final results in Section~\ref{sec:main-results} by showing when the improvements emerge during adaptation.

\section{Conclusion}
\label{sec:conclusion}

We examined latent audio generation from the decoder's perspective and identified a train-generation mismatch between codec training and inference. This mismatch shows that strong reconstruction performance does not necessarily transfer to generation, making the decoder an important but underexplored target for improving latent audio generative models. Based on this observation, AudioGAR constructs intermediate latents between reconstruction and generation and uses them for generation-aware codec decoder adaptation, leaving the codec encoder and latent generative model unchanged. More broadly, our results suggest that improving two-stage generative systems does not always require modifying the generator itself: adapting the decoder to its generation-time inputs provides a complementary route that preserves the pretrained generative structure.

\printbibliography

\clearpage
\begin{appendices}
\crefalias{section}{appsec}
\crefalias{subsection}{appsec}
\crefalias{subsubsection}{appsec}

\setcounter{equation}{0}
\renewcommand{\theequation}{\thesection.\arabic{equation}}

\Huge{Appendix}

\small
\section{Latent Fréchet Distance}
\label{app:latent-fd}

To measure the distributional discrepancy in latent space, we adopt the Fréchet distance between two sets of latent variables, $z_i^{a} \sim \mathcal{P}_A$ and $z_i^{b} \sim \mathcal{P}_B$. Specifically, we estimate the empirical mean and covariance of the two latent distributions, denoted by $(\mu_1, \Sigma_1)$ and $(\mu_2, \Sigma_2)$, respectively. Based on these statistics, we define the latent Fréchet distance (Latent-FD) as
\begin{equation}
\mathrm{FD}(\mathcal{P}_A, \mathcal{P}_B) 
= \|\mu_1 - \mu_2\|_2^2 
+ \mathrm{Tr}\!\left( \Sigma_1 + \Sigma_2 - 2(\Sigma_1 \Sigma_2)^{1/2} \right),
\end{equation}
This metric provides a well-established measure of discrepancy between two distributions by capturing both first-order (mean) and second-order (covariance) statistics. It is closely related to the widely used Fréchet Inception Distance (FID)~\citep{Heusel2017GANsTB} for image generation and Fréchet Distance (FD) and Fréchet Audio Distance (FAD) for audio generation. All of these metrics compute the Fréchet distance under a Gaussian assumption and differ primarily in the representation space used to estimate the corresponding statistics. Specifically, FID uses 2048-dimensional features extracted by a pretrained InceptionV3~\citep{Szegedy2016RethinkingTI} network, FD uses PANNs embeddings~\citep{Kong2019PANNsLP}, and FAD uses VGGish embeddings~\citep{Hershey2016CNNAF}. In contrast, our Latent-FD is computed directly in the model's latent space using latents obtained from the encoder or the generative process. This allows Latent-FD to quantify distributional discrepancies within the generative pipeline while retaining the same underlying geometric interpretation as FID, FD, and FAD.

\section{Decoder Feature Spaces for Latent Distribution Analysis}
\label{app:decoder space}

In Section~\ref{sec:distribution-mismatch}, we analyze intermediate representations of AudioX~\citep{Tian2026AudioXTurboAU} and Stable Audio Open~\citep{Evans2025StableAO}, which use the same codec decoder~\citep{Evans2025StableAO}. We extract features from three decoder stages: (i) \emph{Conv In}, the initial convolutional projection of the latent input; (ii) \emph{Block 2}, an early decoder stage; and (iii) \emph{Block 4}, a later decoder stage.

For encoder-induced and generator-produced latents, we collect activations using forward hooks at \path{pretransform.model.decoder.layers.0}, \path{pretransform.model.decoder.layers.2.layers.4}, and \path{pretransform.model.decoder.layers.4.layers.4}, corresponding to \emph{Conv In}, \emph{Block 2}, and \emph{Block 4}, respectively. The activations are flattened and projected onto a one-dimensional LDA direction fitted to distinguish encoder-induced from generator-produced representations.

\section{Additional LDA analyses on Stable Audio Open}
\label{app:additional-lda}

\begin{wrapfigure}[19]{r}{0.63\textwidth}
\small
\vspace{-9mm}
\begin{center}
\includegraphics[width=0.63\textwidth]{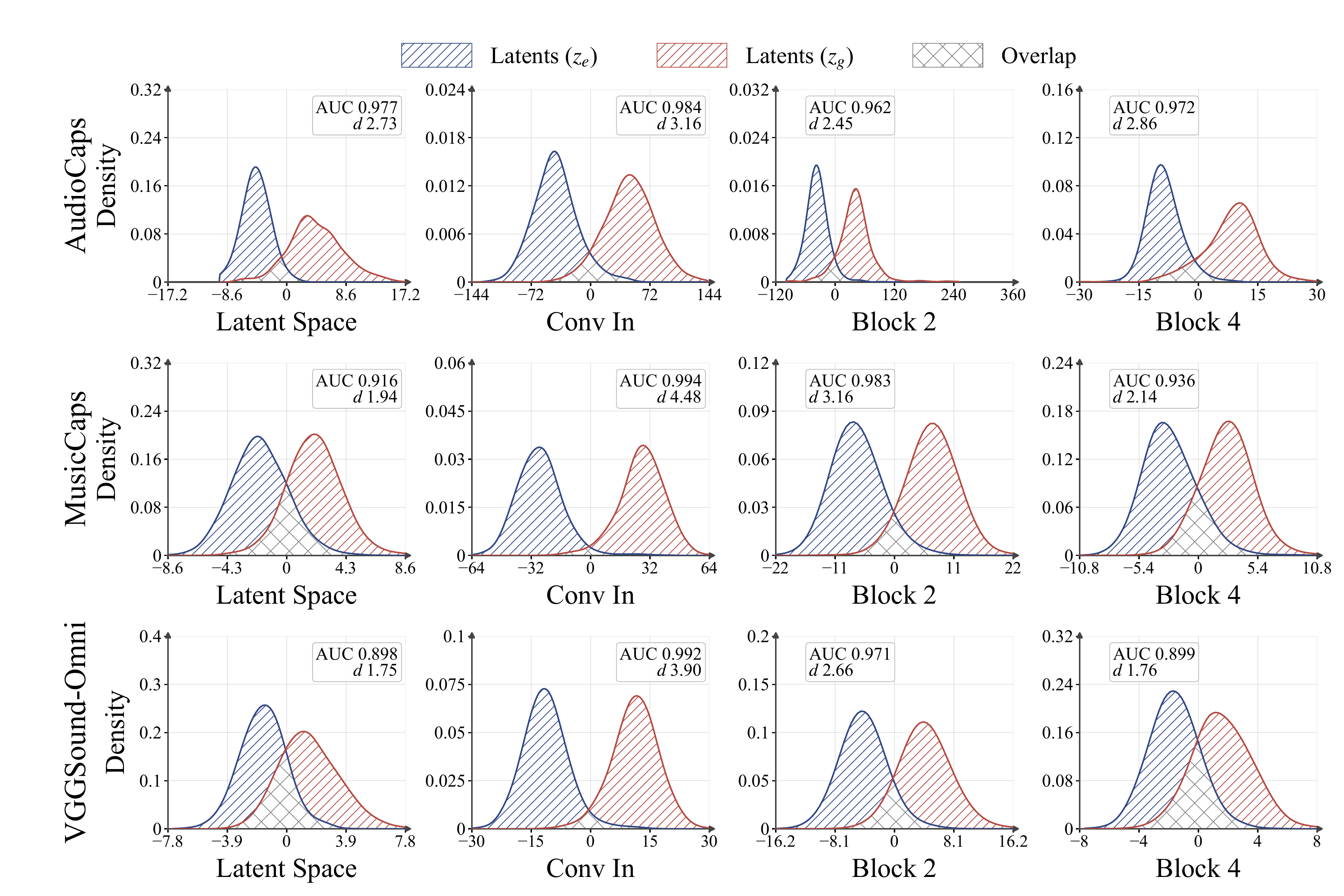}
\vspace{-8mm}
\caption{\small{LDA projections of $z_e$ and $z_g$ in Stable Audio Open. Rows correspond to AudioCaps, MusicCaps, and VGGSound-Omni datasets; columns show latent space and decoder features at Conv In, Block 2, and Block 4. Curves show Gaussian-smoothed, normalized histogram densities of the one-dimensional LDA scores.}}
\label{fig:lda-sao}
\end{center}
\vspace{-2mm}
\end{wrapfigure}
In addition to the LDA analysis of AudioX in Section~\ref{sec:distribution-mismatch}, we conduct the same analysis on Stable Audio Open~\citep{Evans2025StableAO} to examine whether the observed distribution mismatch extends to another latent audio generative model. As shown in Figure~\ref{fig:lda-sao}, encoder-induced and generator-produced latents remain clearly distinguishable across all three datasets. In the latent space, the LDA AUC is 0.977 on AudioCaps~\citep{Kim2019AudioCapsGC}, 0.916 on MusicCaps~\citep{Agostinelli2023MusicLMGM}, and 0.898 on VGGSound-Omni~\citep{Dai2026Omni2SoundTU}. Clear separation between $z_e$ and $z_g$ also persists in the decoder feature spaces at Conv In, Block 2, and Block 4. These results provide additional empirical evidence that $\mathcal{P}_e \neq \mathcal{P}_g$ across different latent audio generative models and datasets.

\section{Additional Quantile Analysis}
\label{app:additional-quantile}
\begin{figure*}[!t]
    \vspace{-2mm}
	\centering
    \captionsetup[subfloat]{skip=-0.5pt}
	\subfloat[MusicCaps]{ 
		\label{fig:fd-quantile-level-musiccaps}  
		\includegraphics[width=0.485\textwidth]{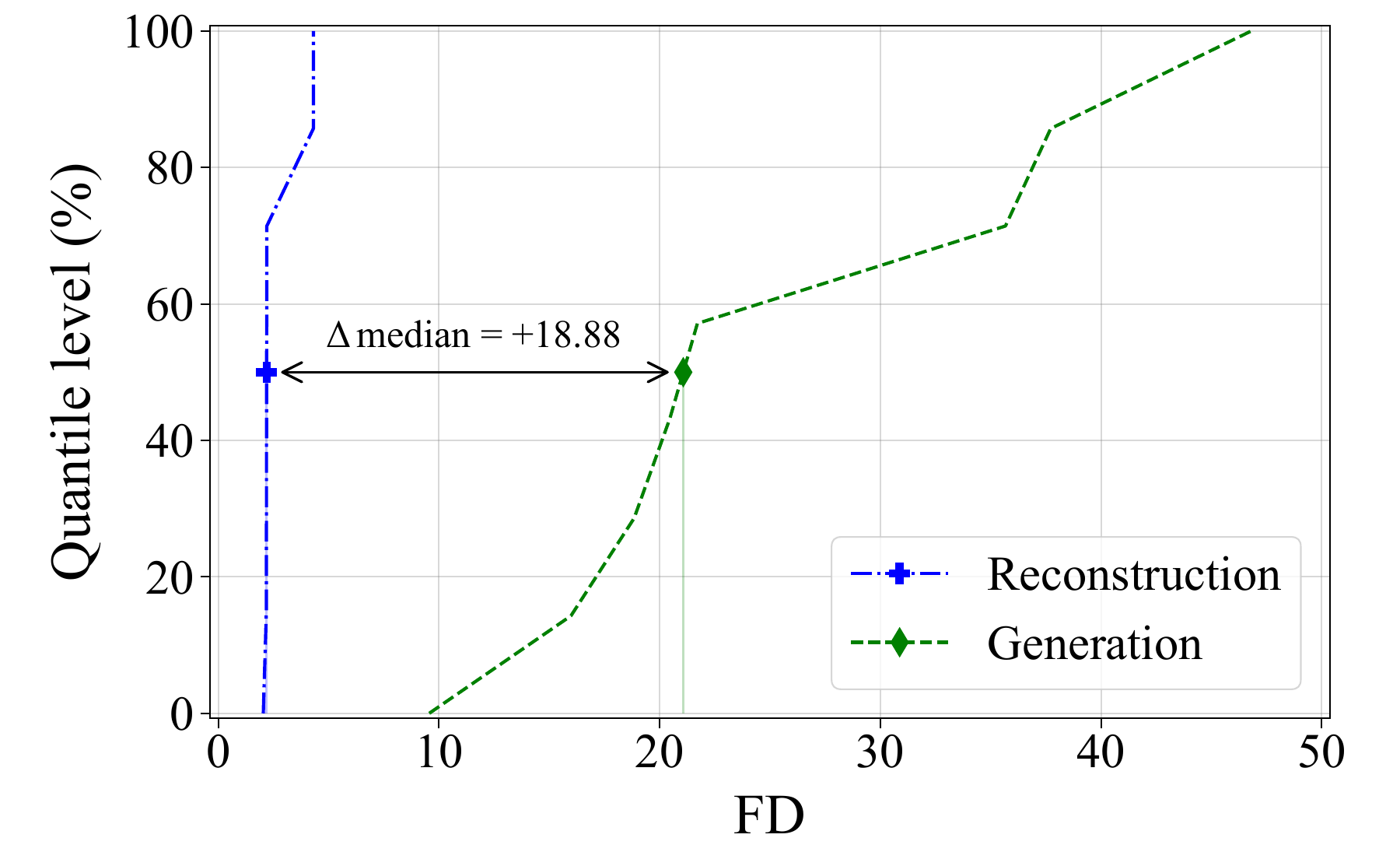}
    }
	\subfloat[VGGSound-Omni]{
		\label{fig:fd-quantile-level-vggsound-omni}
		\includegraphics[width=0.485\textwidth]{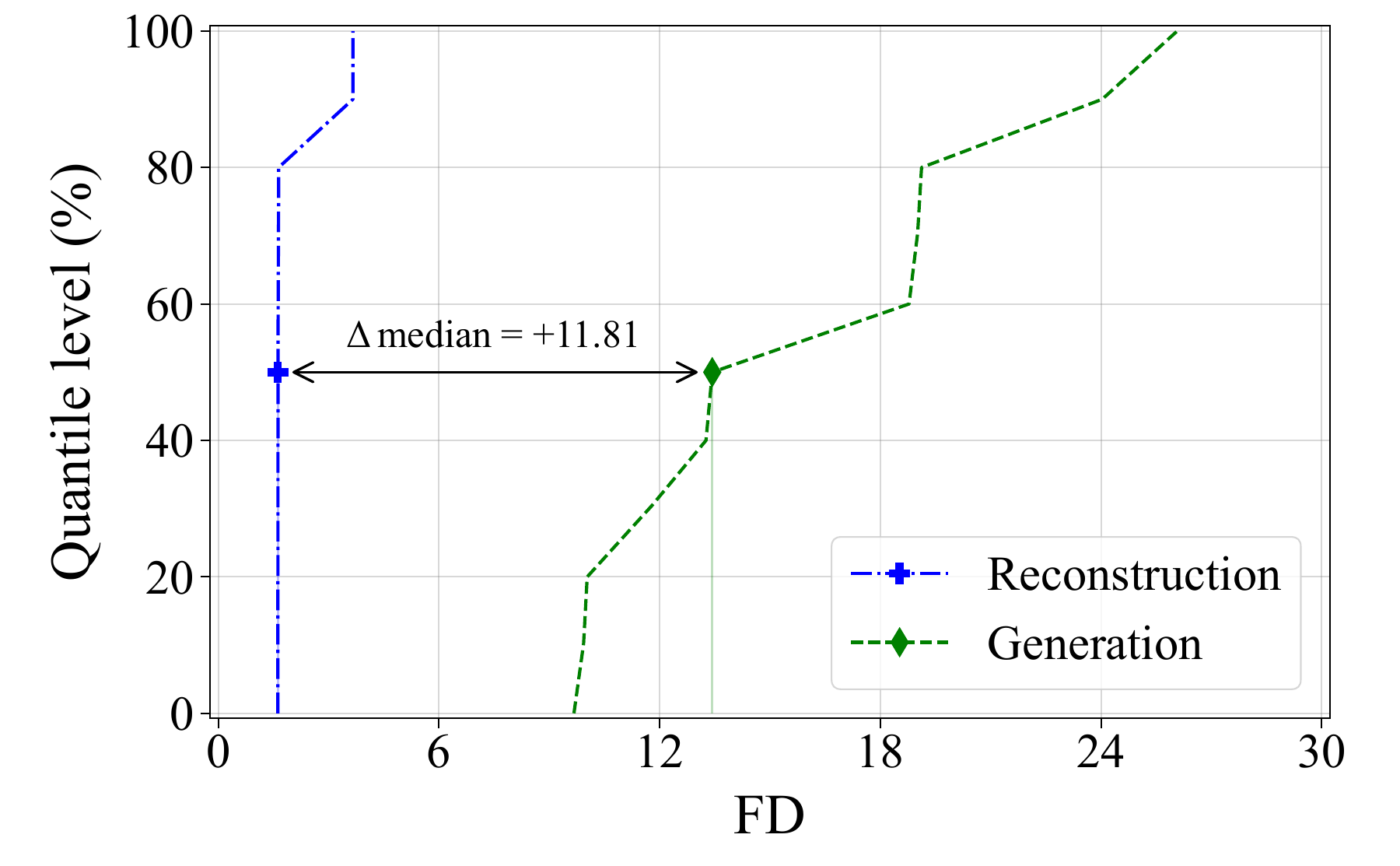}
	}
    \vspace{-2ex}
    \caption{\small{\textbf{Quantile comparison of rFD and gFD on MusicCaps and VGGSound-Omni.} The rFD and gFD curves are separated across the full quantile range on both datasets, with median gFD-rFD gaps of 18.88 on MusicCaps and 11.81 on VGGSound-Omni.}}
    \label{fig:fd-quantile-level}
    \vspace{-2ex}
\end{figure*}
In addition to the quantile analysis of AudioCaps in Section~\ref{sec:recon-gen-gap}, we compare the empirical quantiles of rFD and gFD on MusicCaps~\citep{Agostinelli2023MusicLMGM} and VGGSound-Omni~\citep{Dai2026Omni2SoundTU}. As shown in Figures~\ref{fig:fd-quantile-level}, the rFD and gFD curves remain separated across the full quantile range on both datasets. The median gFD-rFD gap is 18.88 on MusicCaps and 11.81 on VGGSound-Omni. Together with the AudioCaps results, the quantile analysis shows a similar reconstruction-generation gap across all three datasets.

\section{Experimental Details}
\label{app:experimental-details}

\paragraph{Training Data.} AudioX is originally trained on 33,052.28 hours of audio collected from multiple datasets \\ ~\citep{Tian2026AudioXTurboAU}. In contrast, our decoder adaptation uses only 508.3 hours of training audio, corresponding to approximately 1.54\% of the original AudioX training hours. As summarized in Table~\ref{tab:training-data-audiox}, we sample data from four text-to-audio datasets used by AudioX: AudioCaps~\citep{Kim2019AudioCapsGC}, WavCaps~\citep{Mei2024WavCapsAC}, IF-caps~\citep{Tian2026AudioXTurboAU}, and AudioTime~\citep{Xie2025AudioTimeTA}. The resulting subset contains 182,392 samples, with approximately ten seconds per example. All waveforms are resampled to 44.1~kHz and represented in stereo. Samples are drawn with replacement across the four datasets, and duplicated training examples may therefore occur.

\begin{table*}[h]
\vspace{-1ex}
\caption{\small{\textbf{Comparison of the original AudioX training data and the data used for our decoder adaptation.} AudioX training statistics are reported in Appendix A.1 of~\citep{Tian2026AudioXTurboAU}. Our adaptation uses only four text-to-audio datasets and 508.3 hours of audio in total.}}
\vspace{-2ex}
\label{tab:training-data-audiox}
\centering
\resizebox{0.85\textwidth}{!}{%
\begin{tabular}{lrrrrr}
\hline
\multirow{2}{*}{\textbf{Dataset}}  & \multicolumn{2}{c}{\textbf{AudioX}}
& \multicolumn{3}{c}{\textbf{Ours}} \\\cmidrule(lr){2-3}\cmidrule(lr){4-6}
& \textbf{Clips} & \textbf{Hours}
& \textbf{Clips} & \textbf{Share} & \textbf{Hours} \\
\hline
AudioCaps & 45k & 125.1 & 67554 & 36.9\% & 187.4 \\
WavCaps & 108.3k & 300.8 & 46866 & 25.6\% & 130.0 \\
IF-caps & 1268k & 3,524.4 & 58324 & 32.9\% & 167.1 \\
AudioTime & 20k & 355.5 & 9648 & 4.7\% & 23.8 \\
Private T2M dataset & 175.2k & 11679.3 & 0 & 0.0\% & 0.0 \\
V2M & 5685.7k & 15793.58 & 0 & 0.0\% & 0.0 \\
MUCaps & 22k & 1273.6 & 0 & 0.0\% & 0.0 \\
\hline
\textbf{Total} & \textbf{7324.2k} & \textbf{33052.28}
& \textbf{182392} & \textbf{100.0\%} & \textbf{508.3} \\
\hline
\end{tabular}}
\end{table*}

\textbf{TangoMusic.} TangoMusic~\citep{Kong2024ImprovingTM} is originally trained on approximately 5,800 hours of AudioSet data \\ ~\citep{Gemmeke2017AudioSA}. In contrast, our decoder adaptation uses a substantially smaller subset containing 644 hours of ten-second recordings, corresponding to approximately 11.1\% of the original training hours. Following TangoMusic's data preparation procedure~\citep{Kong2024ImprovingTM}, Audio Flamingo~\citep{Kong2024AudioFlamingo} generates candidate text captions from the source audio, which are then ranked and filtered using CLAP~\citep{Wu2023LargeSC} audio-text similarity. Up to three captions per recording are retained with CLAP similarity scores of at least 0.35. All waveforms are resampled to 16~kHz and converted to mono.

\paragraph{Data and preprocessing.} AudioX uses the codec from Stable Audio Open~\citep{Evans2025StableAO}. Accordingly, we follow the preprocessing used for Stable Audio Open codec training. We retain approximately the first ten seconds of each recording and normalize each segment to $-16$ LUFS (Loudness Units relative to Full Scale), with a uniformly sampled loudness jitter of $\pm2$~dB. These segments are then used to construct the AudioGAR latents $z_g^t$ for decoder adaptation. The latent cache is generated with classifier-free guidance (CFG)~\citep{Ho2021ClassifierFreeDG} using a guidance scale of 7.0, matching the CFG setting used for evaluation in AudioX~\citep{Tian2026AudioXTurboAU}.

TangoMusic uses the codec from AudioLDM~\citep{Liu2023AudioLDMTG}. We therefore follow the corresponding waveform preprocessing and acoustic frontend, together with the HiFi-GAN training procedure~\citep{Kong2020HiFiGAN}. Each waveform is normalized by subtracting its mean and scaling its peak amplitude to half of the full-scale amplitude. We then use the first ten seconds of each recording to construct the AudioGAR latents $z_g^t$, which are subsequently decoded by the VAE decoder. The latent cache is generated with CFG using a guidance scale of $3.0$, matching the setting used for TangoMusic evaluation.

\paragraph{Training Setup.} For AudioX, we adapt only the codec decoder while keeping the codec encoder and latent generative model frozen. We follow the Stable Audio Open codec training recipe~\citep{Evans2025StableAO}, using the same GAN training procedure, loss functions, and scalar loss weights. Training is performed for 100k steps with AdamW, using learning rates of $1\times10^{-5}$ for the decoder and $3\times10^{-4}$ for the discriminator, with an effective batch size of 16 clips. Following the original codec training setup, optimization uses aligned 1.5-second audio crops. Training is conducted on a single NVIDIA H100 80GB GPU and takes approximately 10.3 hours.

For TangoMusic, we jointly adapt the VAE decoder and HiFi-GAN vocoder while keeping the VAE encoder and latent generative model frozen. We follow the HiFi-GAN training procedure and objective~\citep{Kong2020HiFiGAN}. Training is performed for 30k steps with AdamW, using a learning rate of $4\times10^{-5}$ for both the VAE decoder and vocoder and $1\times10^{-4}$ for the discriminator, with a batch size of eight clips. Vocoder optimization uses random 0.52-second segments aligned to latent frames, following the HiFi-GAN training procedure~\citep{Kong2020HiFiGAN}. Training is conducted on a single NVIDIA H100 80GB GPU and takes approximately 4.8 hours.

\section{Additional Details on Evaluation Metrics}
\label{app:metrics}

Following AudioX~\citep{Tian2026AudioXTurboAU}, we use the following metrics to evaluate audio and music generation performance.

\textbf{Fréchet Audio Distance (FAD).} FAD measures the distributional discrepancy between generated and real audio in a learned audio feature space, and is analogous to FID~\citep{Heusel2017GANsTB} in image generation. Specifically, it compares the statistics of embedding distributions extracted from generated and reference audio, with a lower value indicating a smaller discrepancy between the two distributions. We compute FAD using features extracted by VGGish~\citep{Hershey2016CNNAF}.

\textbf{Fréchet Distance (FD).} FD follows the same distribution-level comparison as FAD but uses PANNs~\citep{Kong2019PANNsLP} embeddings instead of VGGish, providing a complementary view in a different feature space.

\textbf{Kullback-Leibler Divergence (KL).} KL divergence measures the discrepancy between the acoustic-content distributions of generated and reference audio. Following prior work, we obtain multi-label class predictions from a pretrained PANNs model and compute the divergence between the corresponding prediction distributions. Lower KL indicates greater similarity between generated and reference audio in terms of the predicted acoustic content.

\textbf{Inception Score (IS).} IS evaluates generated samples based on both quality and diversity, favoring confident per-sample class predictions while encouraging variety across the generated set~\citep{Salimans2016ImprovedTF}. Following prior audio-generation work~\citep{Donahue2019AdversarialAS,Majumder2024Tango2A,Liu2023AudioLDMTG}, we use a pretrained PANNs model~\citep{Kong2019PANNsLP} to obtain the class-prediction distributions used to compute IS.

\textbf{Production Complexity (PC) and Production Quality (PQ).} PC and PQ are introduced in the Meta Audiobox Aesthetics framework~\citep{Tjandra2025MetaAA} as no-reference measures of audio aesthetics. PQ captures technical production quality such as clarity and frequency balance; PC captures scene complexity, reflecting the presence of multiple sound components. Unlike distribution-level metrics such as FD and FAD, both metrics can be evaluated independently for each audio clip without requiring a reference sample.

\section{Limitations}
\label{app:limitation}

AudioGAR relies on a denoising trajectory through the frozen latent diffusion model, and is therefore directly applicable to latent diffusion and flow-matching generative models. It does not directly extend to autoregressive latent generators such as AudioGen~\citep{Kreuk2023AudioGenTG} and MusicGen~\citep{Copet2023SimpleAC}, which do not provide the same intermediate denoising trajectory.

The noise level $\eta_t$ controls the trade-off between source correspondence and generation relevance, and intermediate values generally work best in our experiments. However, the preferred $\eta_t$ may depend on the generative model, codec architecture, and dataset. We currently select it empirically rather than with a principled rule.

Our AudioGAR adaptation experiments focus on AudioX and TangoMusic for text-to-audio and text-to-music generation, although we observe the decoder train-generation mismatch across a broader set of models and datasets. It remains to be tested whether the same adaptation strategy transfers to additional latent generative models, codec architectures, and conditioning settings such as video-to-audio and image-to-audio generation. We also evaluate with automatic metrics only and do not include perceptual listening studies.

\end{appendices}


\end{document}